\documentclass[review]{elsarticle}
\usepackage{graphicx}
\usepackage{subfig}
\usepackage{ulem}
\usepackage{amsmath}
\usepackage{xcolor}
\usepackage{hyperref}
\usepackage[capitalise]{cleveref}
\usepackage[top=1.1in, bottom=1.1in, left=1.0in, right=1.0in]{geometry}
\usepackage{lineno,hyperref}
\modulolinenumbers[5]
\usepackage{bm}
\usepackage{graphicx}
\usepackage{graphics}
\usepackage{amsmath}
\usepackage{multirow}
\usepackage{booktabs}
\usepackage{cases}
\usepackage{tabularx}
\begin{document}
\title{Modified steady discrete unified gas kinetic scheme for multiscale radiative heat transfer}
\author[add1]{Xinliang Song}
\author[add1]{Yue Zhang}
\author[add3]{Xiafeng Zhou\corref{cor1}}
\ead{zhouxiafeng@hust.edu.cn}
\author[add4]{Chuang Zhang}
\author[add2]{Zhaoli Guo\corref{cor1}}
\ead{zlguo@mail.hust.edu.cn}
\cortext[cor1]{Corresponding author}
  \address[add1]{State Key Laboratory of Coal Combustion, School of Energy and Power Engineering, Huazhong University of Science and Technology, Wuhan 430074, China}
  \address[add2]{Institute of Interdisciplinary Research for Mathematics and Applied Science, Huazhong University of Science and Technology, Wuhan 430074, China}
  \address[add3]{Department of Nuclear Engineering and Technology, School of Energy and Power Engineering, Huazhong University of Science and Technology, Wuhan 430074, China}
  \address[add4]{Department of Mechanics and Aerospace Engineering, Southern University of Science and Technology, Shenzhen 518055, China}

\begin{abstract}
In this work, a steady discrete unified gas kinetic scheme (SDUGKS) is proposed to solve the steady radiative transfer equation (RTE),
which is an improvement of the original SDUGKS [X. F. Zhou \textit{et al.}, J. Comput. Phys.  423, 109767 (2020)].
The trapezoidal rule other than the rectangular rule used in the original SDUGKS is adopted in the proposed method in the reconstruction of energy flux across cell interface, just as the unsteady DUGKS.
By this way, the characteristic line length of the modified SDUGKS establishes a relationship with the Courant-Friedrichs-Lewy (CFL) number in the DUGKS, which guarantees the accuracy of the modified SDUGKS. Furthermore, the characteristic line length is no longer limited by the extinction coefficient like in original SDUGKS.
As a result, the modified SDUGKS is more accurate and robust than original SDUGKS, and more efficient than the DUGKS for steady radiation problems.
Furthermore, the smooth linear interpolation and the van Leer limiter are used for problems with smooth and discontinuous optical thicknesses, respectively.
Several numerical tests with optical thickness varying from optical thin to thick are conducted to validate the present scheme.
Numerical results demonstrate that the modified SDUGKS can serve as an effective tool in the study of multiscale steady radiative heat transfer in participating media.
\end{abstract}
\maketitle

\section{Introduction}\label{sec1}

Radiative heat transfer appears in many research fields and engineering applications such as combustion systems~\cite{machado2020experimental,gunnarsson2020full}, solar radiation~\cite{hunter2014normalization,fuqiang2017radiative}, nuclear reactor physics~\cite{chahlafi2012radiative}, atmospheric radiation~\cite{liou2002introduction,huang2022effects}, radiative hypersonic flows~\cite{broc1998nonequilibrium,shang2012nonequilibrium}, fluorescence imaging~\cite{gorpas2010three} and other processes.
Most of these problems involve a wide range of optical thicknesses.
Since the velocity of photon is about $3 \times 10^{8}$ m/s, it is reasonable to ignore the transient component and only consider the steady behaviors~\cite{zhang2007nano,modest2013radiative}.
Traditionally, the diffusion equation (DE) which describes the radiative heat transfer at the macroscopic scale is an efficient way for optical thick problems, but it becomes invalid for problems involving optical thin regions~\cite{roger2014hybrid}.
On the contrary, the photon transport equation, or radiative transfer equation (RTE), can be employed to describe radiative process for all optical thickness. However, as a high dimensional integral-differential equation, it is a challenging task to develop efficient numerical schemes with the capability of solving the RTE in all regimes.

In the past few decades, a number of numerical methods have been developed to solve the RTE, such as Monte Carlo method (MCM)~\cite{howell1969application,rubinstein2016simulation}, discrete ordinate method (DOM)~\cite{truelove1987discrete,zhou2019modified}, finite element method (FEM)~\cite{grissa2007three}, finite volume method (FVM)~\cite{raithby1990finite,raithby1999discussion}, spectral element method~\cite{zhao2007solution}, zonal method~\cite{hottel1958radiant}, discrete transfer method (DTM)~\cite{henson1997comparison,sarvari2017solution}, and DRESOR method~\cite{zhou2004new,huang2013solution}. Several recent reviews are also available~\cite{modest2013radiative,kulacki2018handbook,howell2020thermal}.
Among those methods, the MCM, as a powerful and robust probabilistic approach, has achieved great success. However, the method is computationally expensive for participating media with large optical thickness where a great number of photon bundles need to be traced to decrease the statistic errors~\cite{wang2019quantitative}.
Similarly, the deterministic DOM, FEM and FVM also become inefficient when encountering optical thick media, because the cell size of these methods is also limited by the photon mean free path (MFP).
For instance, the diamond difference (DD) method, a popular DOM, usually suffers from nonphysical oscillations in optical thick media unless the mesh is fine enough.
In order to solve multiscale radiative transfer problems accurately and efficiently, some robust methods have been proposed.
A common strategy is using the domain decomposition technique, in which the domain is decomposed into mesoscopic and macroscopic subdomains~\cite{gorpas2010three,golse2003domain}. However, such method faces a critical problem in finding the coupling rule between the mesoscopic and the macroscopic subdomains.
Usually, a buffer zone between two neighboring decomposed domains is adopted~\cite{roger2013dynamic}. Unfortunately, the solution of the coupled system in the buffer zone may be not easy~\cite{coelho2016multi}.
An alternative method for multiscale problem is to solve the RTE in a unified approach, such as the unified gas kinetic scheme (UGKS)~\cite{xu2010unified,zhu2021first} and the discrete unified gas kinetic scheme (DUGKS)~\cite{guo2013discrete,guo2021progress}.
Comparing with the domain decomposition method, the unified methods avoid the coupling of different subdomains.
By considering the photon transport and collision simultaneously, these two methods have been applied efficiently and accurately to a series of transient multiscale radiative transfer problems~\cite{mieussens2013asymptotic,sun2015asymptotic,luo2018multiscale,song2020discrete,shi2021improved} without requiring the cell size smaller than the photon MFP.
However, both UGKS and DUGKS are mainly designed for transient problems.
Recently, the DUGKS has been extended to solve steady multiscale neutron transfer problems~\cite{zhou2020discrete}, and the numerical results showed that the steady DUGKS (SDUGKS) can achieve a good acceleration compared with the DUGKS for steady multiscale neutron transfer problems.
However, a rectangular rule was used in the SDUGKS to reconstruct the interface angular flux. As a result, the method may be not accurate enough especially for optical thick problem with coarse mesh. In addition, the key parameter characteristic line length is strictly restricted by the cell size and the extinction coefficient.

In this work, a modified steady discrete unified gas kinetic scheme is proposed for steady multiscale radiative heat transfer problems. Similar to the DUGKS, a trapezoidal rule is used in the reconstruction of interface angular flux. Therefore, the characteristic line length establishes a subtle connection with the Courant-Friedrichs-Lewy (CFL) number in DUGKS to ensure the same accuracy.
Consequently, the characteristic line length of the present scheme is only limited by the cell size and no longer restricted by the extinction coefficient.
In order to distinguish the two kinds of steady DUGKS, the scheme developed in~\cite{zhou2020discrete} is remarked as the R-SDUGKS, and the scheme proposed in the present work is denoted as T-SDUGKS.

The remainder of this paper is organized as follows. Section~\ref{sec2} introduces the steady gray radiative transfer equation. The T-SDUGKS for steady gray radiative transfer equation is given in Section~\ref{sec3}. Numerical results and discussions are presented in Section~\ref{sec4}. Finally, Section~\ref{sec5} gives a brief summary.

\section{Steady gray radiative transfer equation}\label{sec2}

The steady gray radiative transfer equation can be written as~\cite{modest2013radiative}
\begin{equation}
    \bm s \cdot \nabla I \left( \bm x , \bm s \right) = -\beta \left( \bm x \right)  I \left( \bm x , \bm s \right)+ \beta \left( \bm x \right) S  \left( \bm x , \bm s \right),
\label{eq:RTE}
\end{equation}
\begin{equation}
    S  \left( \bm x , \bm s \right) = \left(1- \omega \left( \bm x \right) \right) I_b \left( \bm x \right)+ \frac{\omega \left( \bm x \right) }{4 \pi} \int_{4\pi} I(\bm x,\bm s ')  d {\bm{\Omega}'},
\label{eq:RTEsource}
\end{equation}
where $I \left( \bm x , \bm s \right)$ is the radiation intensity at position $\bm x $ and direction $ \bm s $, $ \beta \left( \bm x \right) $ is the extinction coefficient related to the local photon mean free path ($\lambda$) as $\lambda=1/ \beta$, $S  \left( \bm x , \bm s \right)$ is the source term of the RTE, $\omega \left( \bm x \right)$ is the scattering albedo, $ I_b \left( \bm x \right) $ is the blackbody intensity, and $\bm{\Omega}'$ is the corresponding solid angle.
For equilibrium problems, the blackbody intensity can be calculated according to energy conservation
\begin{equation}
    I_b \left( \bm x \right)=\frac{1}{4 \pi} \int_{4\pi} I(\bm x,\bm s)  d {\bm{\Omega}}.
\label{eq:Ibequil}
\end{equation}
For nonequilibrium problems with a given temperature field of the medium, the blackbody intensity can be calculated by the Stefan-Boltzmann law~\cite{modest2013radiative},
\begin{equation}
    I_{b}\left( \bm x \right)=\frac{\sigma T^{4}\left( \bm x \right)}{\pi},
\label{eq:Ibnonequil}
\end{equation}
where $\sigma$ is the Stefan-Boltzmann constant and $T\left( \bm x \right)$ is the local temperature of the medium.

In radiative heat transfer problems, the local incident radiation energy $G $ and heat flux $ \bm q $ can be determined by,
\begin{equation}
    G \left( \bm x \right) = \int_{4 \pi} I \left( \bm x , \bm s   \right) d \bm{\Omega} ,
\label{eq:G}
\end{equation}
\begin{equation}
    \bm q \left( \bm x \right)= \int_{4 \pi} I \left( \bm x , \bm s  \right) \bm s d \bm{\Omega} .
\label{eq:q}
\end{equation}

\section{T-SDUGKS for steady radiative transfer equation}\label{sec3}
\subsection{ Discretization of steady radiative transfer equation}\label{sec31}
In order to numerically solve the RTE, Eqs.~\eqref{eq:RTE} and~\eqref{eq:RTEsource} are first discretized in physical and solid angle spaces. The continuous physical space is divided into a set of cells with the finite volume scheme. The solid angle is discretized into $M$ discrete angles using certain quadrature rules, such as Gauss-Legendre quadrature, Gauss-Chebyshev quadrature, and $S_{N}$ quadrature. Then the corresponding discrete forms can be written as
\begin{equation}
    \frac{1}{V_{j}} \sum_{f} \left( {\bm s}_{k} \cdot {\bm n}_{f} \right) I \left( {\bm x}_{f} , {\bm s}_{k}  \right) \Delta S_{f}=
    -\beta \left( {\bm x}_{j} \right) I \left( {\bm x}_{j} , {\bm s}_{k}  \right) +\beta \left( {\bm x}_{j} \right) S \left( {\bm x}_{j} , {\bm s}_{k}  \right),
\label{eq:disRTE}
\end{equation}
\begin{equation}
    S \left( {\bm x}_{j} , {\bm s}_{k}  \right)=\left( 1-\omega \left( {\bm x}_{j} \right) \right) I_{b} \left( {\bm x}_{j} \right)+ \frac{\omega \left( {\bm x}_{j} \right) }{4 \pi} \sum_{m=1}^{M} I \left( {\bm x}_{j} , {\bm s}_{m} ' \right) \omega_{m},
\label{eq:disRTEsourcee}
\end{equation}
where $I \left( {\bm x}_{j} , {\bm s}_{k}  \right)$ denotes the cell averaged radiation intensity in the control volume $V_{j}$ centered at ${\bm x}_{j}$ along the photon propagation direction ${\bm s}_{k}$, ${\bm n}_{f}$ is the outward unit normal vector at the interface ${\bm x}_{f}$,  $\Delta S_{f}$ is the corresponding interface area, and $\omega_{m}$ and ${\bm s}_{m} ' $ are the weight and discrete angle of the corresponding spherical quadrature, respectively.

\subsection{Reconstruction of the interface radiation intensity}\label{sec32}
To update the radiation intensity $I \left( {\bm x}_{j} , {\bm s}_{k}  \right)$, the reconstruction of the interface intensity $I \left( {\bm x}_{f} , {\bm s}_{k}  \right)$ becomes the key point.
By integrating Eq.~\eqref{eq:disRTE} along the direction ${\bm s}_{k}$ with a certain length $l$ and applying the trapezoidal rule to the right hand term, one can obtain
\begin{equation}
\begin{split}
    &\frac{I \left( {\bm x}_{f} , {\bm s}_{k}  \right)-I \left( {\bm x}_{f} - l {\bm s}_{k} , {\bm s}_{k}  \right)}{l}= \\
    &\frac{\beta({\bm x}_{f})}{2} \left[ S \left( {\bm x}_{f} , {\bm s}_{k}  \right)- I \left( {\bm x}_{f} , {\bm s}_{k}  \right)  \right] + \frac{\beta \left({\bm x}_{f} -l {\bm s}_{k} \right)}{2} \left[ S \left( {\bm x}_{f} - l {\bm s}_{k} , {\bm s}_{k} \right) -  I \left( {\bm x}_{f} - l {\bm s}_{k} , {\bm s}_{k} \right) \right],
\label{eq:TrapeFace}
\end{split}
\end{equation}
where both the transport and collision terms are coupled simultaneously. We note that the rectangular rule is used in the R-SDUGKS~\cite{zhou2020discrete}.
By reviewing the interface radiation intensity in transient radiative heat transfer, the relationship between the characteristic line length in Eq.~\eqref{eq:TrapeFace} and the CFL used in Ref.~\cite{luo2018multiscale} can be easily established as
\begin{equation}
    l=\frac{1}{2} \alpha \Delta x_{min},
\label{eq:charaLf}
\end{equation}
where $0 < \alpha <1$ is the CFL number and $\Delta x_{min}$ is the minimal mesh size. It is obvious that the characteristic line length is no longer limited by the maximal extinction coefficient as mentioned in Ref.~\cite{zhou2020discrete}.

In order to remove the implicit term of Eq.~\eqref{eq:TrapeFace}, two new distribution functions are introduced,
\begin{equation}
    {\bar{I}}^{+} \left( \bm x , \bm s  \right) = I \left( \bm x , \bm s  \right)- \frac{\beta \left(\bm x \right) l}{2} \left( I \left( \bm x , \bm s  \right) - S \left( \bm x , \bm s  \right) \right),
\label{eq:Distr1}
\end{equation}
\begin{equation}
    {\bar{I}} \left( \bm x , \bm s  \right) = I \left( \bm x , \bm s  \right) + \frac{\beta \left(\bm x \right) l}{2} \left( I \left( \bm x , \bm s  \right) - S \left( \bm x , \bm s  \right) \right).
\label{eq:Distr2}
\end{equation}
Substituting Eqs.~\eqref{eq:Distr1} and~\eqref{eq:Distr2} into Eq.~\eqref{eq:TrapeFace}, one can obtain
\begin{equation}
    {\bar{I}} \left( {\bm x}_{f} , {\bm s}_{k} \right) = {\bar{I}}^{+} \left( {\bm x}_{f} - l {\bm s}_{k} , {\bm s}_{k}  \right).
\label{eq:substi}
\end{equation}
For smooth problems, ${\bar{I}}^{+} \left( {\bm x}_{f} - l {\bm s}_{k} , {\bm s}_{k}  \right)$ can be approximated  as
\begin{equation}
     {\bar{I}}^{+} \left( {\bm x}_{f} - l {\bm s}_{k} , {\bm s}_{k}  \right)=   {\bar{I}}^{+} \left( {\bm x}_{f} , {\bm s}_{k}  \right)-l {\bm s}_{k} \cdot \bm{\sigma}_{f},
\label{eq:Smooth}
\end{equation}
\begin{figure}[!ht]
    \centering
    \includegraphics[width=0.4\textwidth]{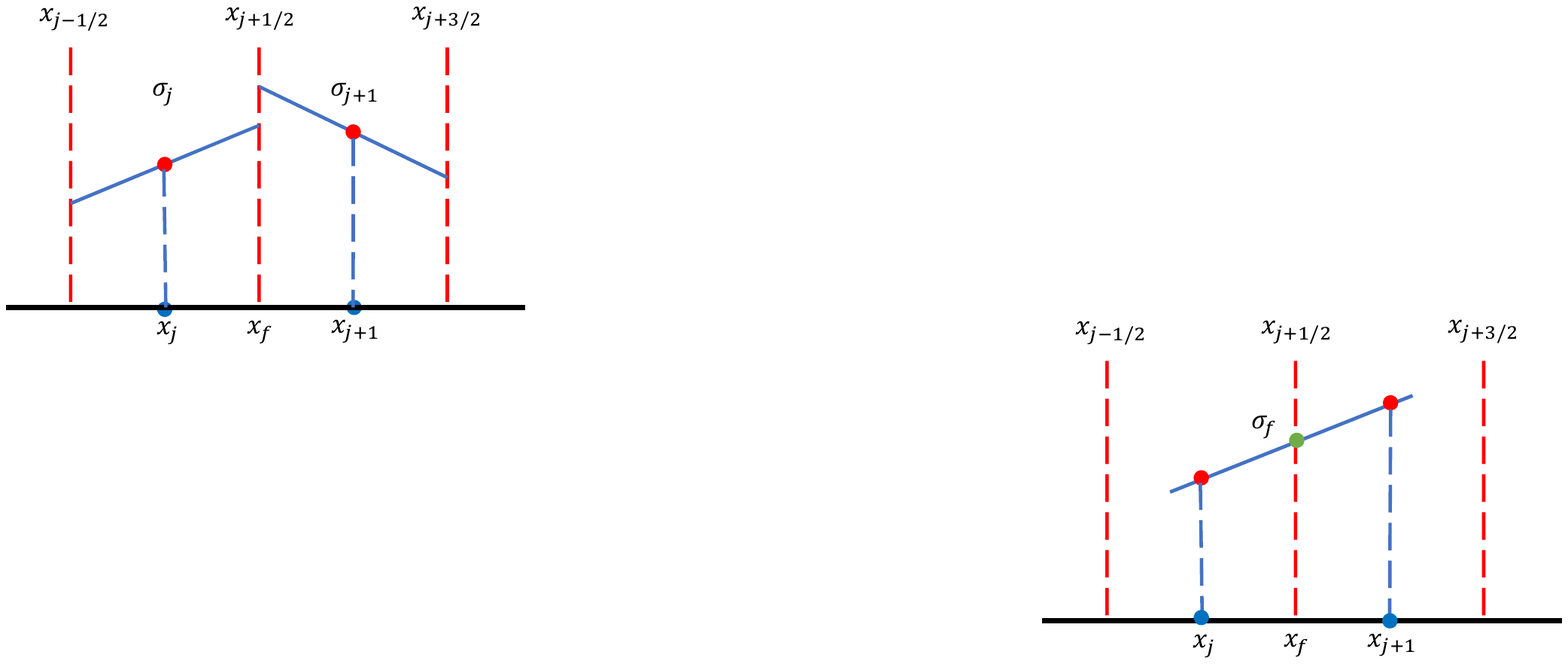}
    \centering
    \caption{Schematic of one-dimensional cell geometry for smooth problems.}
\label{tu0one}
\end{figure}
where ${\bar{I}}^{+} \left( {\bm x}_{f} , {\bm s}_{k}  \right)$ and the gradient $\bm{\sigma}_{f}= \nabla {\bar{I}}^{+} \left( {\bm x}_{f} , {\bm s}_{k}  \right) $ at the cell interface can be approximated by smooth linear interpolation~\cite{guo2013discrete}. As an example, figure~\ref{tu0one} shows the one-dimensional (1D) case. In this case, the reconstructions become
\begin{equation}
     \sigma_{j+1/2}=\frac{{\bar{I}}^{+} \left(  x_{j+1} , \mu_{k}  \right)-{\bar{I}}^{+} \left( x_{j} , \mu_{k}  \right)}{x_{j+1}-x_{j}},
\label{eq:SmSlop}
\end{equation}
\begin{equation}
     {\bar{I}}^{+} \left(  x_{j+1/2} , \mu_{k}  \right)={\bar{I}}^{+} \left(  x_{j} , \mu_{k}  \right)+ \sigma_{j+1/2} \left( x_{j+1/2}-x_{j} \right),
\label{eq:SmIplus}
\end{equation}
where $\mu_{k}$ is the direction cosine of $\bm{s}_{k}$ along $x$ direction.
For discontinuous problems, the van Leer limiter~\cite{van1977towards} can be adopted to reconstruct the gradient in each cell. Assuming that ${\bar{I}}^{+} \left( {\bm x}_{j} , {\bm s}_{k}  \right)$ is linear in cell $V_{j}$, then $ {\bar{I}}^{+} \left( {\bm x}_{f} - l {\bm s}_{k} , {\bm s}_{k}  \right)$ can be reconstructed as
\begin{equation}
    {\bar{I}}^{+} \left( {\bm x}_{f} - l {\bm s}_{k} , {\bm s}_{k}  \right) = {\bar{I}}^{+} \left( {\bm x}_{j} , {\bm s}_{k}  \right) +  \left( {\bm x}_{f} - l {\bm s}_{k}-{\bm x}_{j}  \right)\cdot \bm{\sigma}_{j}, ({\bm x}_{f} - l {\bm s}_{k}) \in V_{j},
\label{eq:recons}
\end{equation}
\begin{figure}[!ht]
    \centering
    \includegraphics[width=0.4\textwidth]{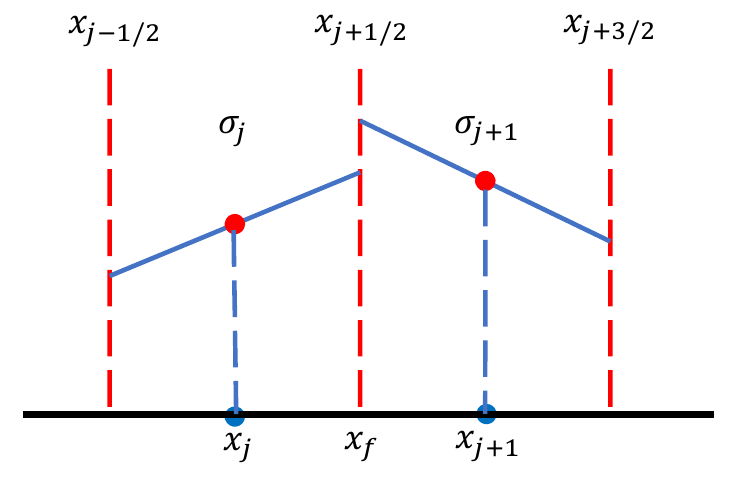}
    \centering
    \caption{Schematic of one-dimensional cell geometry for discontinuous problems.}
\label{tuone}
\end{figure}
where $ \bm{\sigma}_{j}$ is the slope of ${\bar{I}}^{+}$ in cell $j$. For example, in the 1D case shown in Fig.~\ref{tuone}, the variable ${\bar{I}}^{+} \left( x_{f} - l \mu_{k} , \mu_{k}  \right)$ can be approximated as
\begin{equation}
    {\bar{I}}^{+} \left( x_{f} - l \mu_{k} , \mu_{k}  \right)=\left\{
    \begin{array}{rcl}
    {\bar{I}}^{+} \left( x_{j} , \mu_{k} \right)+\left( x_{f} - l\mu_{k} - x_{j}\right) \sigma_{j},      & {\mu_{k} >0}, \\
    {\bar{I}}^{+} \left( x_{j+1} , \mu_{k} \right)+\left( x_{f} - l\mu_{k} - x_{j+1}\right) \sigma_{j+1},           & {\mu_{k} <0}.
    \end{array} \right.
\label{eq:OneDrecons}
\end{equation}
The reconstruction of $\sigma_{j}$ in each cell becomes
\begin{equation}
    \sigma_{j}=\left[ sgn \left(a_{1}\right) +sgn \left(a_{2}\right) \right] \frac{\mid a_{1}\mid \mid a_{2}\mid}{\mid a_{1}\mid + \mid a_{2}\mid},
\label{eq:GetSigma}
\end{equation}
where
\begin{equation}
    a_{1}=\frac{{\bar{I}}^{+} \left( x_{j} , \mu_{k} \right)-{\bar{I}}^{+} \left( x_{j-1} , \mu_{k} \right)}{x_{j}-x_{j-1}},
    a_{2}=\frac{{\bar{I}}^{+} \left( x_{j+1} , \mu_{k} \right)-{\bar{I}}^{+} \left( x_{j} , \mu_{k} \right)}{x_{j+1}-x_{j}}.
\label{eq:a1a2}
\end{equation}
Then ${\bar{I}} \left( {\bm x}_{f} , {\bm s}_{k} \right) $ can be obtained from  Eq.~\eqref{eq:substi}.

Next, integrating Eq.~\eqref{eq:Distr2} over the entire solid angle, one can obtain
\begin{equation}
\int_{4 \pi}  I \left( \bm x , \bm s  \right) d \Omega =
\frac{2}{2+ \beta \left( \bm x \right)  \left( 1- \omega \left( \bm x \right)\right)l}
\int_{4 \pi}  \bar{I} \left( \bm x , \bm s  \right) d \Omega +
\frac{4 \pi \beta \left( \bm x \right)  \left( 1- \omega\left( \bm x \right)\right)l}
{2+ \beta \left( \bm x \right)  \left( 1- \omega\left( \bm x \right)\right)l} I_{b} \left( \bm x \right).
\label{eq:preinterfaceI}
\end{equation}
Then from Eq.~\eqref{eq:Distr2}, the radiation intensity at the cell interface can be computed by
\begin{equation}
\begin{split}
    I \left( {\bm x}_{f} , {\bm s}_{k} \right)
    &= \frac{2}{2+\beta \left( {\bm x}_{f} \right) l}{\bar{I}} \left( {\bm x}_{f} , {\bm s}_{k} \right) +
    \frac{\beta \left( {\bm x}_{f} \right)  \left( 1- \omega\left( {\bm x}_{f} \right)  \right) l }{ 2+ \beta \left( {\bm x}_{f} \right)  \left( 1- \omega\left( {\bm x}_{f} \right)  \right)l } I_{b} \left( {\bm x}_{f} \right)  \\
    &+ \frac{2 \beta \left( {\bm x}_{f} \right) \omega \left( {\bm x}_{f} \right)l}{\left( 2+ \beta \left( {\bm x}_{f} \right) l \right) \left[ 2 + \beta \left( {\bm x}_{f} \right) \left( 1- \omega \left( {\bm x}_{f} \right) \right) l \right]} \frac{1}{4 \pi} \sum_{m=1}^{M} {\bar{I}} \left( {\bm x}_{f} , {\bm s}_{m} ' \right) \omega_{m}.
\label{eq:interfaceI}
\end{split}
\end{equation}

\subsection{Implicit delta formulation}\label{sec33}
With the known interface intensity $I \left( {\bm x}_{f} , {\bm s}_{k} \right)$, the implicit delta formulation method will be used to update the cell center intensity  $I \left( {\bm x}_{j} , {\bm s}_{k} \right)$.
First, Eq.~\eqref{eq:disRTE} is solved using the following iterative scheme
\begin{equation}
    \frac{1}{V_{j}} \sum_{f} \left( {\bm s}_{k} \cdot {\bm n}_{f} \right) I^{n+1} \left( {\bm x}_{f} , {\bm s}_{k}  \right) \Delta S_{f}=
    -\beta \left( {\bm x}_{j} \right) I^{n+1} \left( {\bm x}_{j} , {\bm s}_{k}  \right) +\beta \left( {\bm x}_{j} \right) S^{n} \left( {\bm x}_{j} , {\bm s}_{k}  \right),
\label{eq:disRTEL}
\end{equation}
with
\begin{equation}
    S^{n} \left( {\bm x}_{j} , {\bm s}_{k}  \right)=\left( 1-\omega \left( {\bm x}_{j} \right) \right) I_{b}^{n} \left( {\bm x}_{j} \right)+ \frac{\omega \left( {\bm x}_{j} \right) }{4 \pi} \sum_{m=1}^{M} I^{n} \left( {\bm x}_{j} , {\bm s}_{m} ' \right) \omega_{m}.
\label{eq:disRTEsourceeL}
\end{equation}
where $n$ represents the $n$-th iteration step. If we define the increment between two successive iterations as
\begin{equation}
    \Delta I^{n}\left( \bm x , {\bm s}  \right)=I^{n+1}\left( \bm x , {\bm s}  \right)-I^{n}\left( \bm x , {\bm s}  \right).
\label{eq:increments}
\end{equation}
Then Eq.~\eqref{eq:disRTEL} can be rewritten in the delta form,
\begin{equation}
    \frac{1}{V_{j}} \sum_{f} \left( {\bm s}_{k} \cdot {\bm n}_{f} \right) \Delta I^{n} \left( {\bm x}_{f} , {\bm s}_{k}  \right) \Delta S_{f}+
    \beta \left( {\bm x}_{j} \right) \Delta I^{n} \left( {\bm x}_{j} , {\bm s}_{k}  \right) =  {Res}^{n} \left( {\bm x}_{j} , {\bm s}_{k}  \right),
\label{eq:deltaform}
\end{equation}
where ${Res}^{n} \left( {\bm x}_{j} , {\bm s}_{k}  \right)$ is the residual,
\begin{equation}
    {Res}^{n} \left( {\bm x}_{j} , {\bm s}_{k}  \right) = \beta \left( {\bm x}_{j} \right)S^{n} \left( {\bm x}_{j} , {\bm s}_{k}  \right)-\beta \left( {\bm x}_{j} \right)I^{n} \left( {\bm x}_{j} , {\bm s}_{k}  \right)-\frac{1}{V_{j}} \sum_{f} \left( {\bm s}_{k} \cdot {\bm n}_{f} \right)  I^{n} \left( {\bm x}_{f} , {\bm s}_{k}  \right) \Delta S_{f}.
\label{eq:residual}
\end{equation}
Overall, the cell center radiation intensity $I^{n+1} \left( {\bm x}_{j} , {\bm s}_{k} \right) $ can be obtained from Eqs.~\eqref{eq:increments} and~\eqref{eq:deltaform}. Once the iteration converges, the increment $\Delta I^{n}$ will vanish, suggesting that the two terms on the left hand side of Eq.~\eqref{eq:deltaform} dose not affect the converged solution. Particularly, a first-order upwind scheme will be employed to calculate the first term on the left hand side of Eq.~\eqref{eq:deltaform}. The accuracy of the final results are determined by the right-hand term of Eq.~\eqref{eq:deltaform}, and a higher-order scheme should be employed for this term.
In the present method, the T-SDUGKS is employed to calculate the interface angular flux, i.e., the last term of Eq.~\eqref{eq:residual}.

The one-dimensional case is taken as an example to show the details of solving $ I^{n+1} \left( x_{j} , \mu_{k} \right)$. In this case, Eq.~\eqref{eq:deltaform} can be written as
\begin{equation}
    \frac{1}{V_{j}}  \left( \mu_{k}  \Delta I^{n} \left( x_{j+1/2} , \mu_{k} \right) -  \mu_{k} \Delta I^{n} \left( x_{j-1/2} , \mu_{k} \right) \right)\Delta S_{f} + \beta  \Delta I^{n} \left( x_{j} , \mu_{k} \right) = Res^{n}\left( x_{j} , \mu_{k} \right),
\label{eq:deltaformOneD}
\end{equation}
where the increment of interface intensity is approximated by the upwind scheme, i.e.,
\begin{equation}
    \left\{
    \begin{array}{rcl}
    \Delta I^{n} \left( x_{j+1/2} , \mu_{k} \right)=\Delta I^{n} \left( x_{j} , \mu_{k} \right),      & {\mu_{k} >0}, \\
    \Delta I^{n} \left( x_{j-1/2} , \mu_{k} \right)=\Delta I^{n} \left( x_{j} , \mu_{k} \right),           & {\mu_{k} <0}.
    \end{array} \right.
\label{eq:approximate}
\end{equation}
Substituting Eq.~\eqref{eq:approximate} into Eq.~\eqref{eq:deltaformOneD}, the increment $\Delta I^{n} \left( x_{j} , \mu_{k}  \right)$ can be computed as
\begin{equation}
    \Delta I^{n} \left( x_{j} , \mu_{k}  \right)=\frac{ Res^{n}\left( x_{j} , \mu_{k} \right) V_{j}+ \mid \mu_{k}  \mid \Delta S_{f} \Delta I^{n} \left( x_{in} , \mu_{k} \right)}{\beta \left( x_{j} \right) V_{j} +  \mid \mu_{k} \mid \Delta S_{f} },
\label{eq:GetIncrement}
\end{equation}
where
\begin{equation}
    \Delta I^{n} \left( x_{in} , \mu_{k} \right) = \left\{
    \begin{array}{rcl}
    \Delta I^{n} \left( x_{j-1/2} , \mu_{k} \right),      & {\mu_{k} >0}, \\
    \Delta I^{n} \left( x_{j+1/2} , \mu_{k} \right),           & {\mu_{k} <0}.
    \end{array} \right.
\label{eq:sweeping}
\end{equation}
According to Eqs.~\eqref{eq:increments} and~\eqref{eq:GetIncrement}, the cell center intensity  $ I^{n+1} \left( x_{j} , \mu_{k} \right)$ can be updated. The two- and three-dimensional cases are similar and are not presented here.

\subsection{Boundary condition}\label{sec34}
In this work, the diffusively emitting and reflecting boundaries are considered. The general boundary condition for the steady radiative transfer equation can be expressed as
\begin{equation}
    I \left( {\bm x}_{w} , \bm s_{k} \right)=\epsilon_{w} I_{b} \left( {\bm x}_{w} \right) + \frac{\rho_{w}}{\pi}
    \sum_{{\bm n}_{w} \cdot \bm{s}'_{m} < 0} \left( {\bm n}_{w} \cdot \bm{s}'_{m} \right) I \left( {\bm x}_{w} , \bm{s}'_{m} \right) \omega_{m}, {\bm n}_{w} \cdot \bm{s}_{k} > 0,
\label{eq:boundary}
\end{equation}
where $\epsilon_{w}$ is the diffuse emissivity, $\rho_{w}$ is the diffuse reflectivity, $\epsilon_{w}=1-\rho_{w}$, $I_{b} \left( {\bm x}_{w} \right)$ is the blackbody radiation intensity at the boundary surface with a specified temperature, and ${\bm n}_{w}$ is the inner normal vector at the boundary.

\subsection{Comparison of the numerical fluxes of R-SDUGKS and T-SDUGKS}
We now analyze the differences between the R-SDUGKS and T-SDUGKS.
Firstly, it is noted that the exact solution at the cell interface can be expressed as
\begin{equation}
   I_{exact} \left( {\bm x}_{f} , {\bm s}_{k}  \right)-I \left( {\bm x}_{f} - l {\bm s}_{k} , {\bm s}_{k}  \right)=\int_{0}^{l} A\left( {\bm x}_{f} - l {\bm s}_{k} + h {\bm s}_{k} , {\bm s}_{k}\right) dh,
\label{eq:ExactFace}
\end{equation}
where
\begin{equation}
\begin{split}
   A\left( \bm x, {\bm s}_{k} \right)
  = \frac{\beta \left( \bm x \right) \omega \left( \bm x \right)}{4 \pi}  \sum_{m=1}^{M} I \left( \bm x , {\bm s}_{m} ' \right) \omega_m
   - \beta \left( \bm x \right) I\left( \bm x, {\bm s}_{k} \right)
  +\beta \left( \bm x \right) \left( 1-\omega \left( \bm x \right) \right) I_b \left( \bm x \right).
\label{eq:ExactFaceA}
\end{split}
\end{equation}
If we approximate the integration in Eq.~\eqref{eq:ExactFace} with the rectangular rule,
we will get the R-SDUGKS~\cite{zhou2020discrete}, and Eq.~\eqref{eq:ExactFace} becomes
\begin{equation}
   I_{recta} \left( {\bm x}_{f} , {\bm s}_{k}  \right)-I \left( {\bm x}_{f} - l {\bm s}_{k} , {\bm s}_{k}  \right)=l A\left( {\bm x}_{f} - l {\bm s}_{k} , {\bm s}_{k}\right).
\label{eq:RectanFace}
\end{equation}
On the other hand, the T-SDUGKS will be obtained when the trapezoidal rule is used,
\begin{equation}
   I_{trape} \left( {\bm x}_{f} , {\bm s}_{k}  \right)-I \left( {\bm x}_{f} - l {\bm s}_{k} , {\bm s}_{k}  \right)=\frac{l}{2} \left[ A\left( {\bm x}_{f} - l {\bm s}_{k} , {\bm s}_{k}\right) + A\left( {\bm x}_{f}  , {\bm s}_{k}\right) \right].
\label{eq:TrapezoFace}
\end{equation}
From Eqs.~\eqref{eq:ExactFace},~\eqref{eq:RectanFace}, and~\eqref{eq:TrapezoFace}, we can obtain that
\begin{equation}
   I_{recta} \left( {\bm x}_{f} , {\bm s}_{k}  \right)=  I_{exact} \left( {\bm x}_{f} , {\bm s}_{k}  \right) + O \left( \Delta x^2 \right)   - \frac{l^2}{2} {\bm s}_{k} \cdot \nabla A \left( {\bm x}_{f} , {\bm s}_{k}\right) + O \left( l^3 \right),
\label{eq:rewrRectanFace}
\end{equation}
\begin{equation}
   I_{trape} \left( {\bm x}_{f} , {\bm s}_{k}  \right)=I_{exact} \left( {\bm x}_{f} , {\bm s}_{k}  \right) + O \left( \Delta x^2 \right)+ \frac{l^3}{12} \left( {\bm s}_{k} \cdot \nabla \right) ^2  A \left( {\bm x}_{f} , {\bm s}_{k} \right) + O \left( l^4 \right),
\label{eq:rewrTrapezoFace}
\end{equation}
where the error $O \left( \Delta x^2 \right)$ comes from the spatial approximation of the second term on the left-hand side of Eqs.~\eqref{eq:RectanFace} or~\eqref{eq:TrapezoFace}.
The errors from the quadratures are explicitly expressed in Eqs.~\eqref{eq:rewrRectanFace} or~\eqref{eq:rewrTrapezoFace} for the two schemes.
It can be observed that the approximation error of T-SDUGKS is $O \left( \Delta x^2 \right) + O \left( {l^3} \right)$ while the R-SDUGKS is $O \left( \Delta x^2 \right) + O \left( {l^2} \right)$, which means the T-SDUGKS will be more accurate than the R-SDUGKS  where $O \left( {l^2} \right)$ cannot be ignored.

\subsection{Algorithm}\label{sec35}
In summary, the iteration procedure of the T-SDUGKS can be summarized as follows:
\begin{enumerate}
\item Calculate the interface intensity $I^{n} \left( {\bm x}_{f} , {\bm s}_{k} \right)$.\label{StepA}
    \begin{enumerate}
    \item Calculate ${\bar{I}}^{+,n} \left( {\bm x}_{j} , {\bm s}_{k} \right)$  according to Eq.~\eqref{eq:Distr1}.
    \item Calculate the distribution function $\bar{I}^{n} \left( {\bm x}_{f} , {\bm s}_{k} \right)$ at cell interface according to Eq.~\eqref{eq:substi}.
    \item Calculate the interface intensity $I^{n} \left( {\bm x}_{f} , {\bm s}_{k} \right)$ according to Eq.~\eqref{eq:interfaceI}.
    \end{enumerate}
\item Update the cell center intensity $I^{n+1} \left( {\bm x}_{j} , {\bm s}_{k} \right)$.\label{StepB}
    \begin{enumerate}
    \item Calculate the residual ${Res}^{n} \left( {\bm x}_{j} , {\bm s}_{k} \right)$ according to Eqs.~\eqref{eq:disRTEsourceeL} and~\eqref{eq:residual}. \label{Step1}
    \item Calculate increments $\Delta I^{n} \left( {\bm x}_{j} , {\bm s}_{k} \right)$ according to Eq.~\eqref{eq:deltaform}.\label{Step2}
    \item Update the center intensity $I^{n+1} \left( {\bm x}_{j} , {\bm s}_{k} \right)$ according to Eq.~\eqref{eq:increments}.\label{Step3}
    \end{enumerate}
\item Repeat step~\ref{StepA} - step~\ref{StepB}, until the convergence criteria is satisfied.
\end{enumerate}

When the center intensity is known, the radiation energy $ G\left( {\bm x}_{j} \right) $ and heat flux $ \bm q \left( {\bm x}_{j} \right) $ can be obtained,
\begin{equation}
    G \left( {\bm x}_{j} \right)= \sum_{k=1}^{M} I \left( {\bm x}_{j} , {\bm s}_{k} \right) \omega_{k},
\label{eq:Gdis}
\end{equation}
\begin{equation}
    \bm q \left( {\bm x}_{j} \right)= \sum_{k=1}^{M} I \left( {\bm x}_{j} , {\bm s}_{k} \right) {\bm s}_{k}\omega_{k}.
\label{eq:qdis}
\end{equation}
The convergence criteria in our simulations is defined as
\begin{equation}
    E= \frac{\sum_{j}\mid G^{n+1} \left( {\bm x}_{j} \right) - G^{n} \left( {\bm x}_{j} \right) \mid} {\sum_{j} \mid G^{n+1} \left( {\bm x}_{j} \right) \mid} < 10^{-10}.
\label{eq:criteria}
\end{equation}

\begin{figure}[!b]
    \centering
    \includegraphics[width=0.4\textwidth]{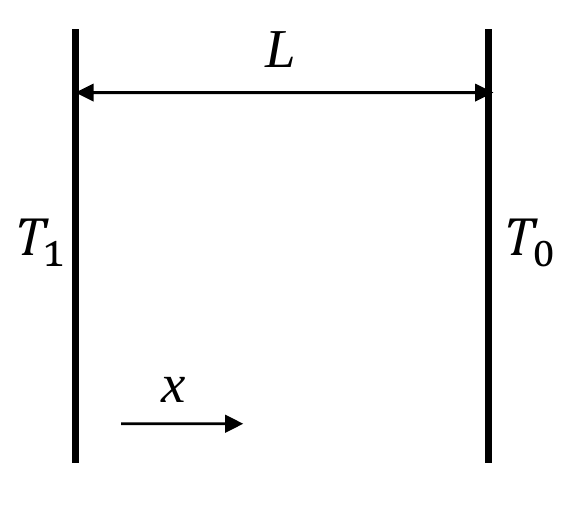}
    \centering
    \caption{Schematic of one-dimensional radiation heat transfer.}
\label{tutwozero}
\end{figure}

\section{Numerical tests}\label{sec4}
In this section, several radiative heat transfer problems with the optical thickness ranging from thin to thick regimes are simulated to validate the numerical properties of the proposed T-SDUGKS. In the numerical implementation, the Gauss-Legendre quadrature~\cite{abramowitz1972handbook} is adopted to discretize the solid angle with the cosine of zenith $\mu \in \left[-1,1\right]$ and the azimuth angle $\varphi \in \left[ 0,2\pi \right]$. The CFL number of transient DUGKS is set to be 0.5. Correspondingly, the characteristic line length in T-SDUGKS is set to be $l =0.25 \Delta x_{min}$ according to Eq.~\eqref{eq:charaLf} unless otherwise stated. All computations are performed on a machine with an Intel(R) Core(TM) i7-8700 CPU @ 3.20GHz processor with 8.00 GB  RAM and 64-bit operating system.
\subsection{Numerical accuracy}
Firstly, some general one- and two-dimensional cases are simulated to prove the accuracy of the present T-SDUGKS. The computation time and steps of the DUGKS, T-SDUGKS and R-SDUGKS are also compared here.
\subsubsection{Radiation in a slab}\label{sec41}
The radiative equilibrium problem in a slab with thickness $L=1$ is shown in Fig.~\ref{tutwozero}. The left wall is kept hot with unity nondimensional emissive power $\left(\Phi_{1}=1 \right)$ while the right wall is kept cold with $\Phi_{0}=0$, where $\Phi=\left( \sigma T^4 - \sigma T_{0}^4 \right)/\left( \sigma T_{1}^4 - \sigma T_{0}^4 \right)$, $T_{0}$ and $T_{1}$ are the temperature of right and left walls, respectively~\cite{heaslet1965radiative}, and both walls are black media. Initially, the participating medium inside the slab is cold. When the slab is filled with pure scattering media $\left( \omega =1 \right)$, the solution of the problem can be expressed as~\cite{heaslet1965radiative},
\begin{equation}
    G^{*} \left( \tau \right)= \frac{1}{2} \left[ E_{2} \left( \tau \right) + \int_{0}^{\tau_L} G^{*} \left( \tau ' \right) E_{1} \left( \mid \tau -\tau ' \mid \right) d \tau ' \right],
\label{eq:analytical}
\end{equation}
where $\tau = \beta x$ is the optical path length, $\tau_{L} = \beta L$ is the optical thickness, and $ E_{n} \left( x \right)=\int_{0}^{1} t^{n-2} exp \left( -x/t \right) dt$ is the exponential integral function.
$G^{*}=\left( G-G_{0} \right)/\left( G_{1}-G_{0} \right)$ is the nondimensional incident radiation energy, where $G_{0}=4 \sigma T_{0}^{4}$ and $G_{1}=4 \sigma T_{1}^{4}$.
\begin{figure}[!ht]
 \centering
\subfloat[]{\includegraphics[width=0.35\textwidth]{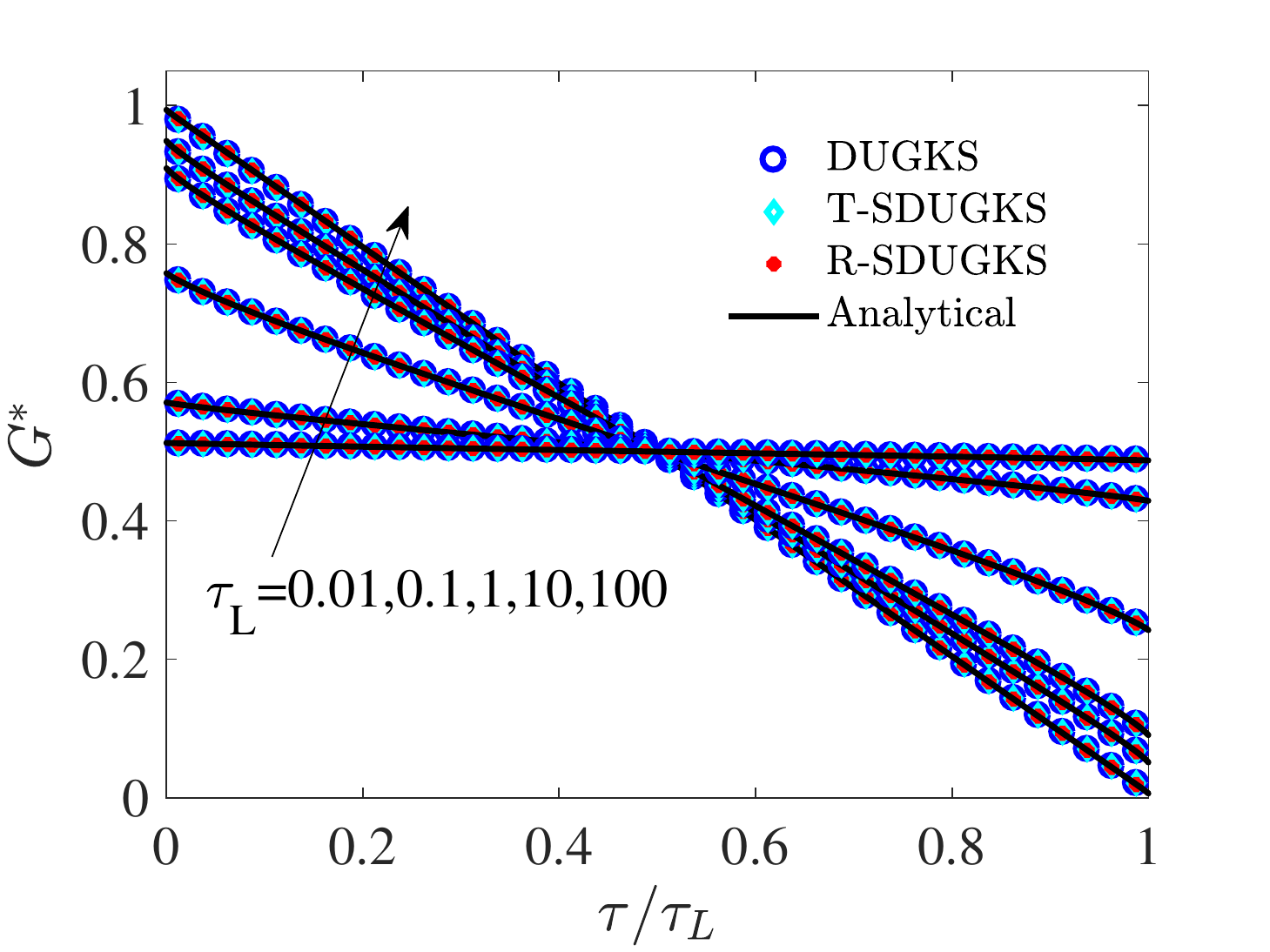}}~~
 \subfloat[]{\includegraphics[width=0.35\textwidth]{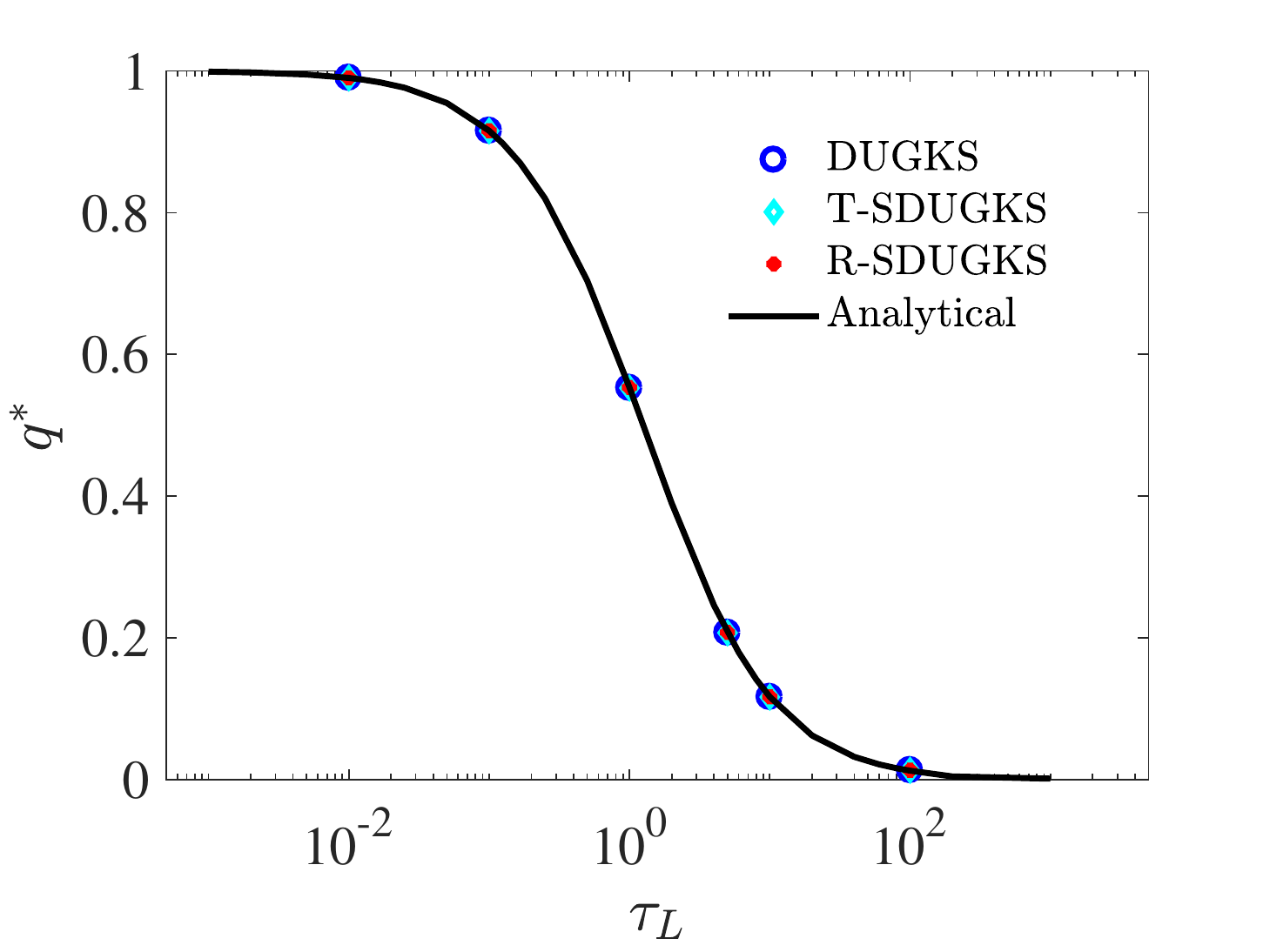}}~~
 \caption{Radiation energy and heat flux profiles with different optical thicknesses.
 }
 \label{figure11}
\end{figure}
In fact, Eq.~\eqref{eq:analytical} can be numerically solved to get a ``numerical exact" solution.
Here we use the trapezoidal quadrature with 4000 points to solve Eq.~\eqref{eq:analytical}, which can give mesh-independent reference solution.
In our numerical simulations the angular space is discretized with the 100-point Gauss-Legendre quadrature, the physical space is discretized into $N_{x}=40$ uniform cells, and the smooth reconstruction is used for the interface radiation intensity.

Figure~\ref{figure11} shows the distribution of nondimensional incident radiation energy $G^{*}$ and heat flux $q^{*}$ with the optical thickness ranging from 0.01 to 100. It can be seen that the results of DUGKS, T-SDUGKS and R-SDUGKS are all in good agreement with the analytical solutions.
The computation time and steps of the three schemes are shown in Table~\ref{tab1}, where ``Rate" represents the speedup ratio in time compared to DUGKS.
It can be observed that the T-SDUGKS and R-SDUGKS are much faster than the transient DUGKS for optical thin problems, but the acceleration rates decrease with increasing optical thickness, which is due to the inherent nature of the source iteration method~\cite{adams2002fast}.
\begin{table}[!h]
\caption{Comparison of the computation costs of different schemes for the slab.}
\centering
\begin{tabular}{c c c c c c c c c c c}
  \hline
  \multirow{2}{*}{$\tau_L$} & \multicolumn{2}{c}{DUGKS}& \multicolumn{3}{c}{T-SDUGKS} & \multicolumn{3}{c}{R-SDUGKS}  \\
 \cline{2-9}
 & Time(s) &  Steps & Time(s) &  Steps &Rate &  Time(s)&  Steps &Rate \\
 \hline
 0.01&   4.962&   61421&  0.01&   98&   496.2& 0.009&  98	&551.3  \\
 0.1&   0.718&   8858&   0.01&	 84&     71.8& 0.01 &  88	& 71.8  \\
 1&     0.196&   2377&   0.009&  67&     21.8& 0.006&  66	& 32.7  \\
 5&     0.289&   3516&   0.027&  277&    10.7& 0.022&  275	& 13.1  \\
 10&    0.444&   5458&   0.07&   791&     6.3& 0.058&  786	&  7.7  \\
 100&   3.061&   37816&  3.898&  46616&   0.8& 3.3  &  46244	&  0.9  \\
 \hline
\end{tabular}
\label{tab1}
\end{table}
\subsubsection{Radiation in a black square enclosure}\label{sec44}
\begin{figure}[!h]
    \centering
    \includegraphics[width=0.4\textwidth]{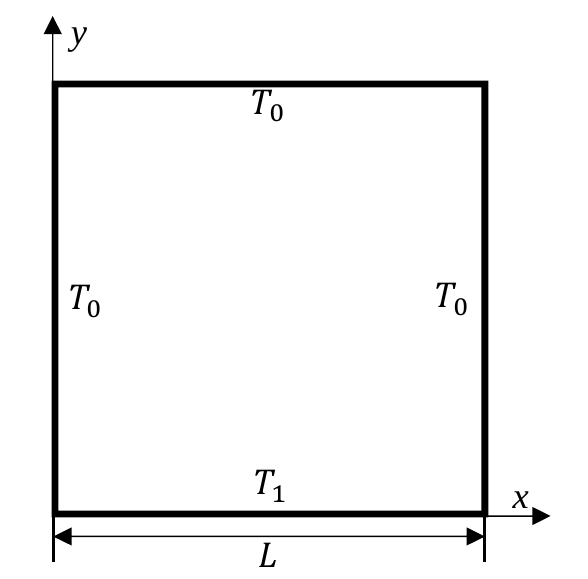}
    \centering
    \caption{Schematic of two-dimensional radiation heat transfer.}
\label{tutwoone}
\end{figure}

In this section, the radiative heat transfer in a 2D black square enclosure of length $L$ filled with pure scattering media is simulated.
The bottom wall is kept hot ($\Phi_{1}=1$), whereas the other walls and the media are kept cold ($\Phi_{0}=0 $). The schematic of the problem is shown in Fig.~\ref{tutwoone}.
\begin{figure}[!ht]
 \centering
 \subfloat[]{\includegraphics[width=0.35\textwidth]{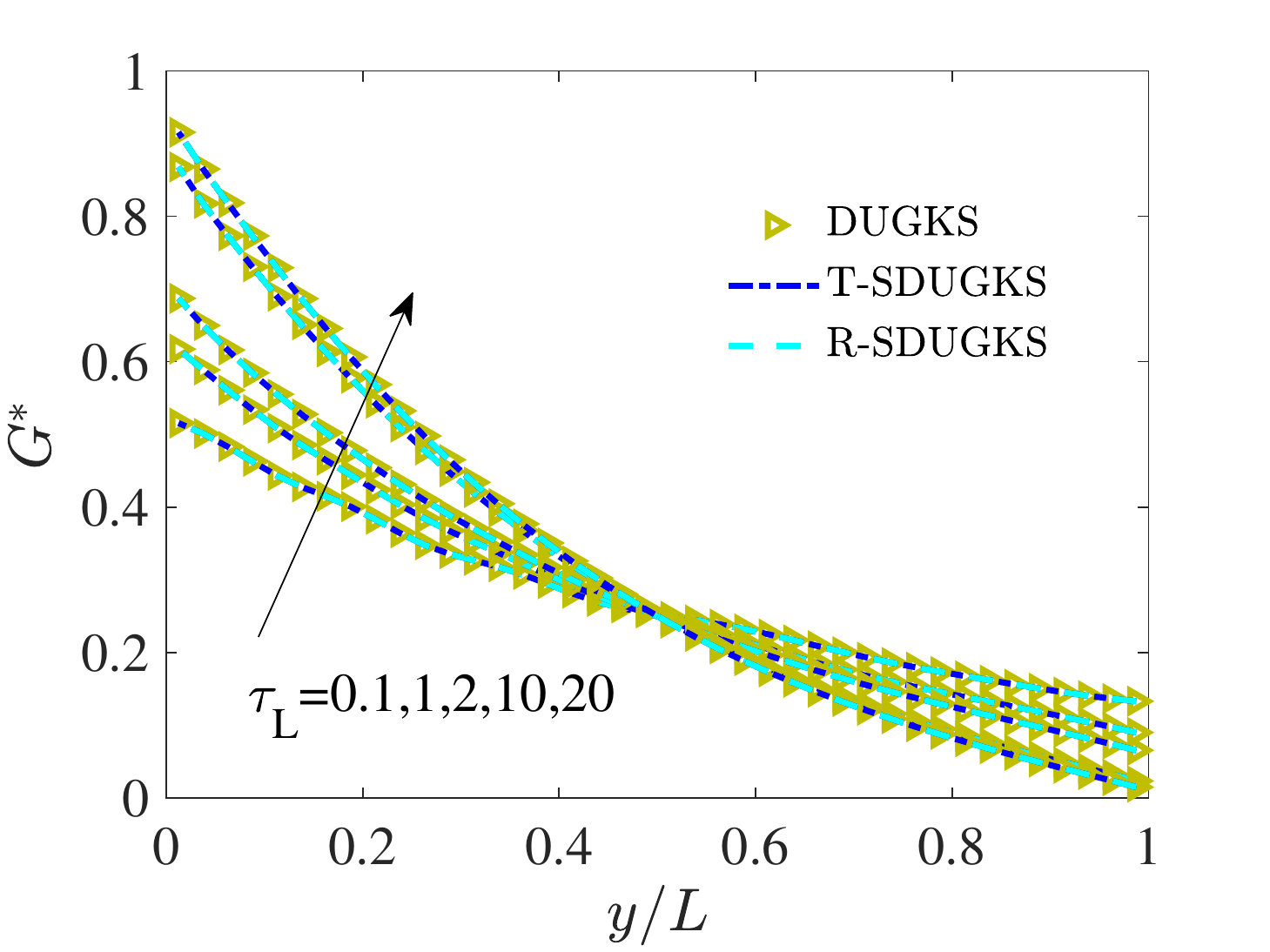}}~~
\subfloat[]{\includegraphics[width=0.35\textwidth]{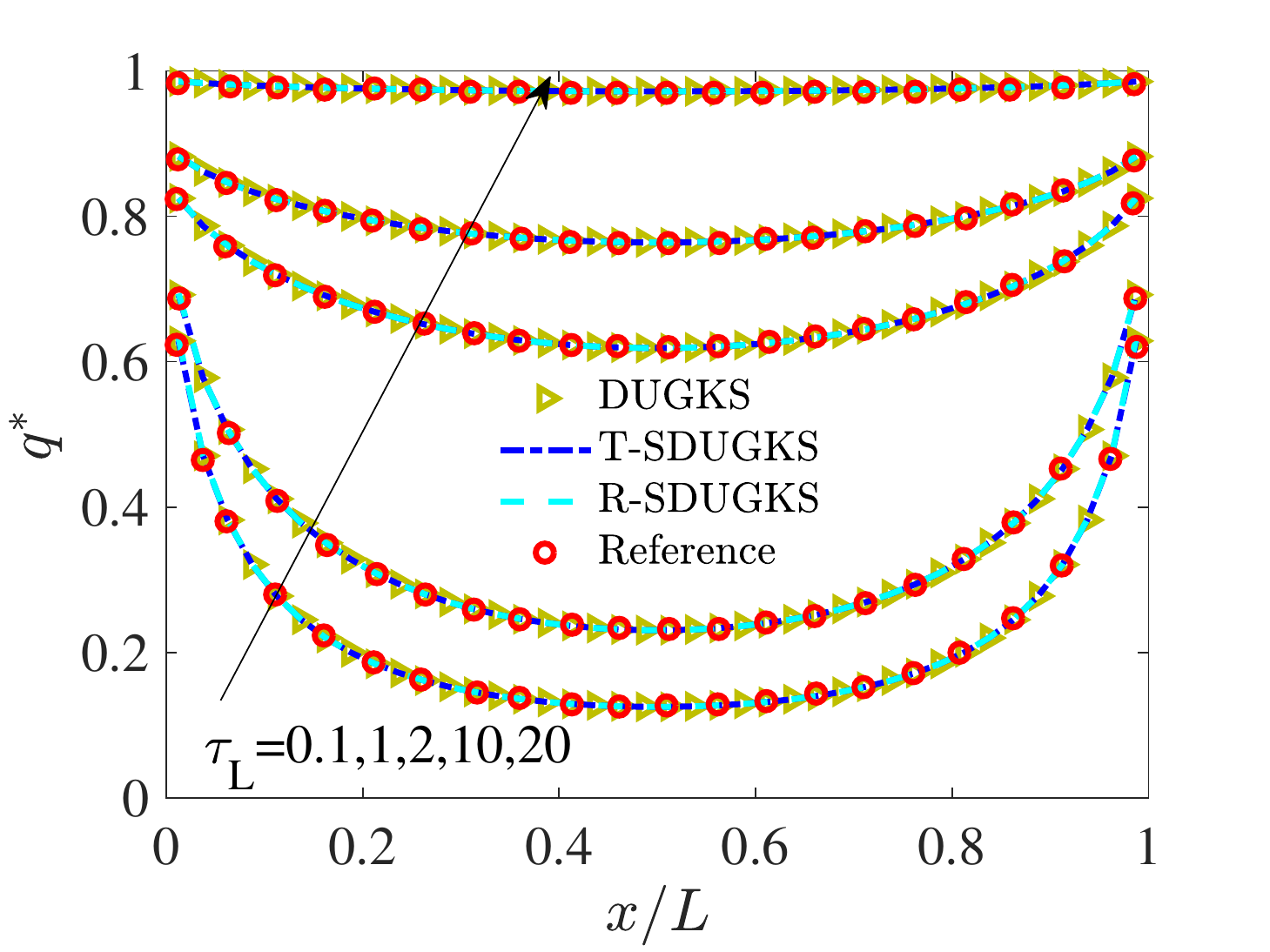}}~~
 \caption{Radiation energy profiles along the vertical centerline and heat flux at the bottom wall with different optical thicknesses.
 }
 \label{figure41}
\end{figure}
\begin{figure}[!ht]
 \centering
\subfloat[]{\includegraphics[width=0.35\textwidth]{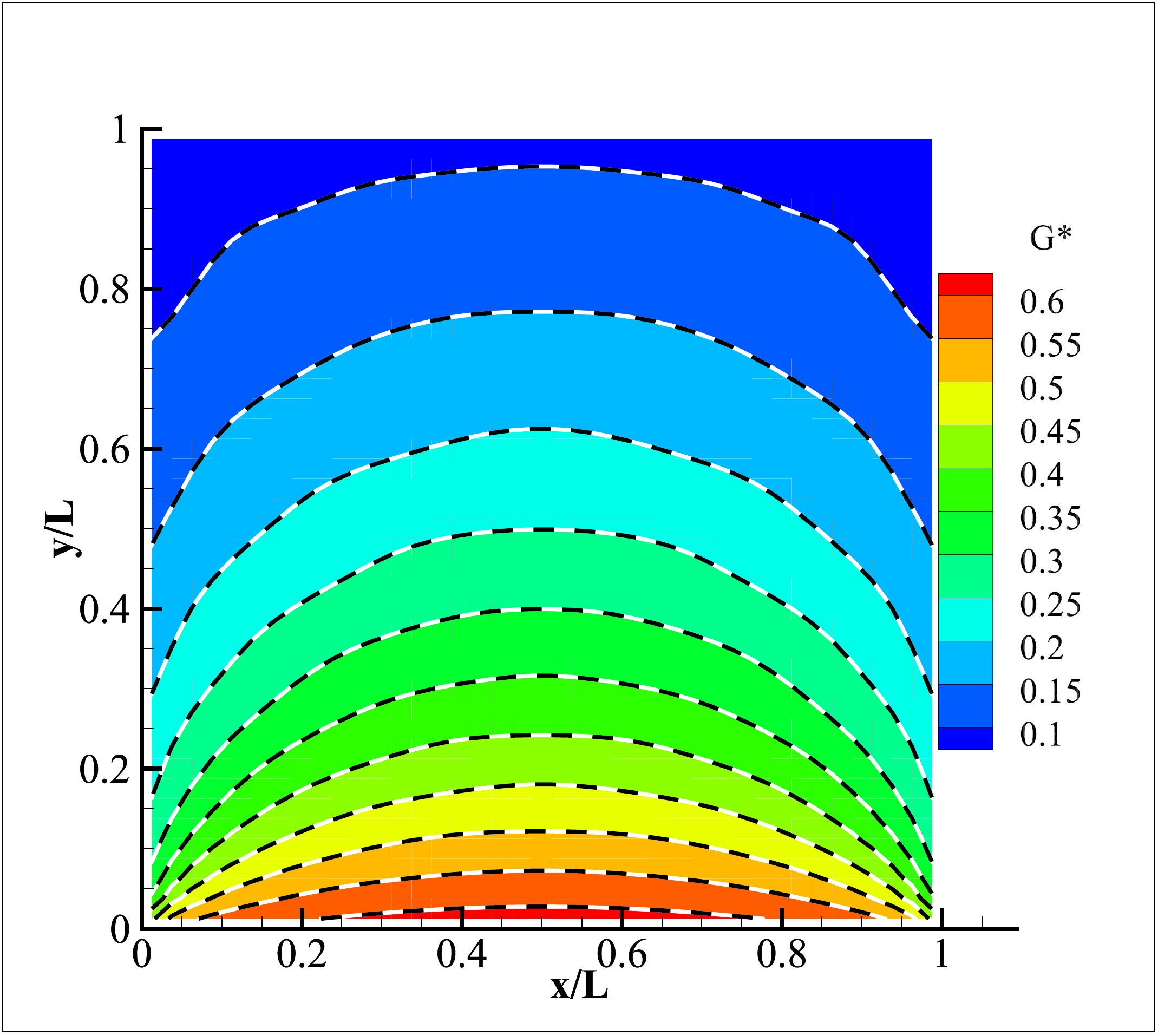}}~~
\subfloat[]{\includegraphics[width=0.35\textwidth]{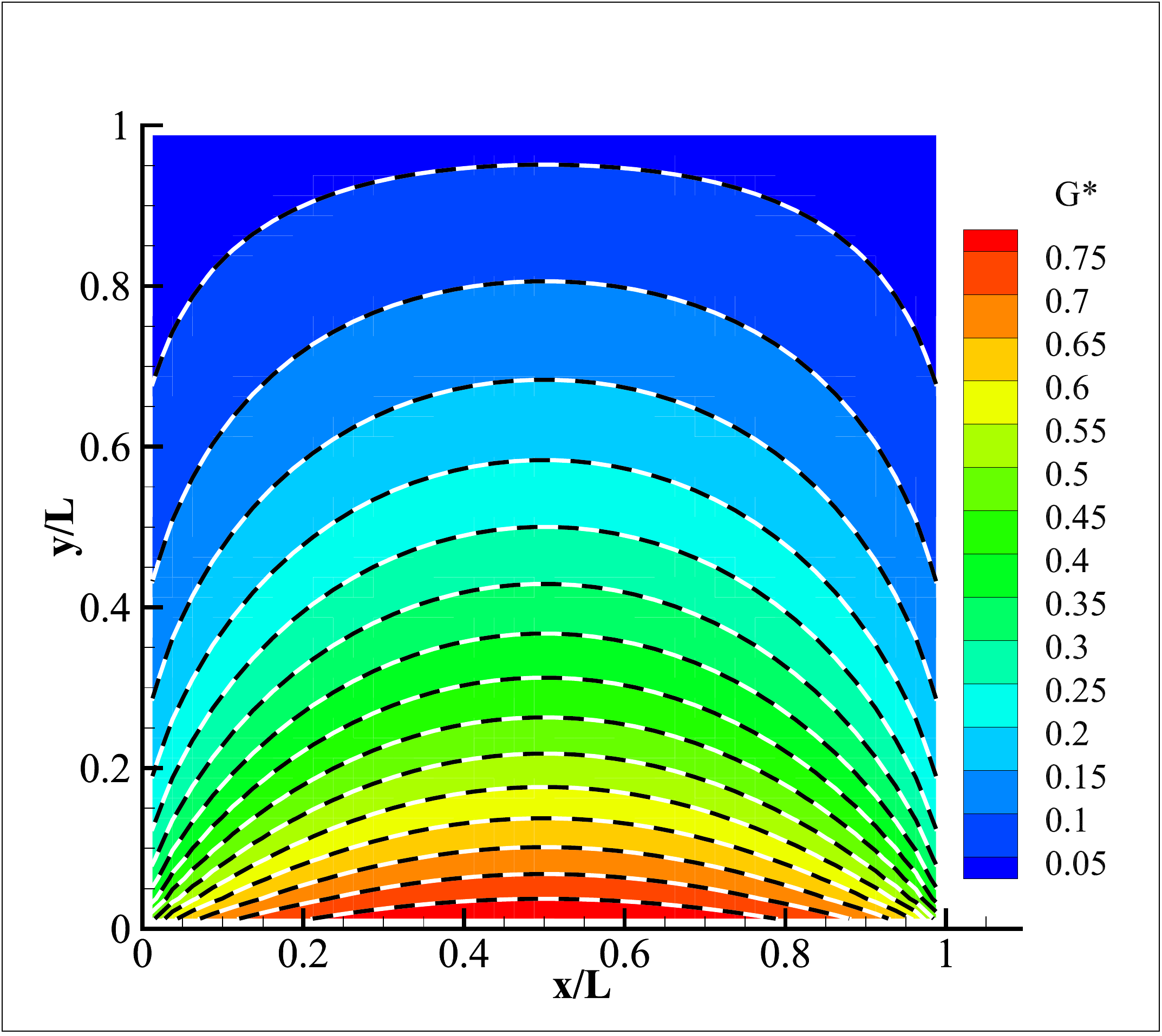}}~~\\
 \subfloat[]{\includegraphics[width=0.35\textwidth]{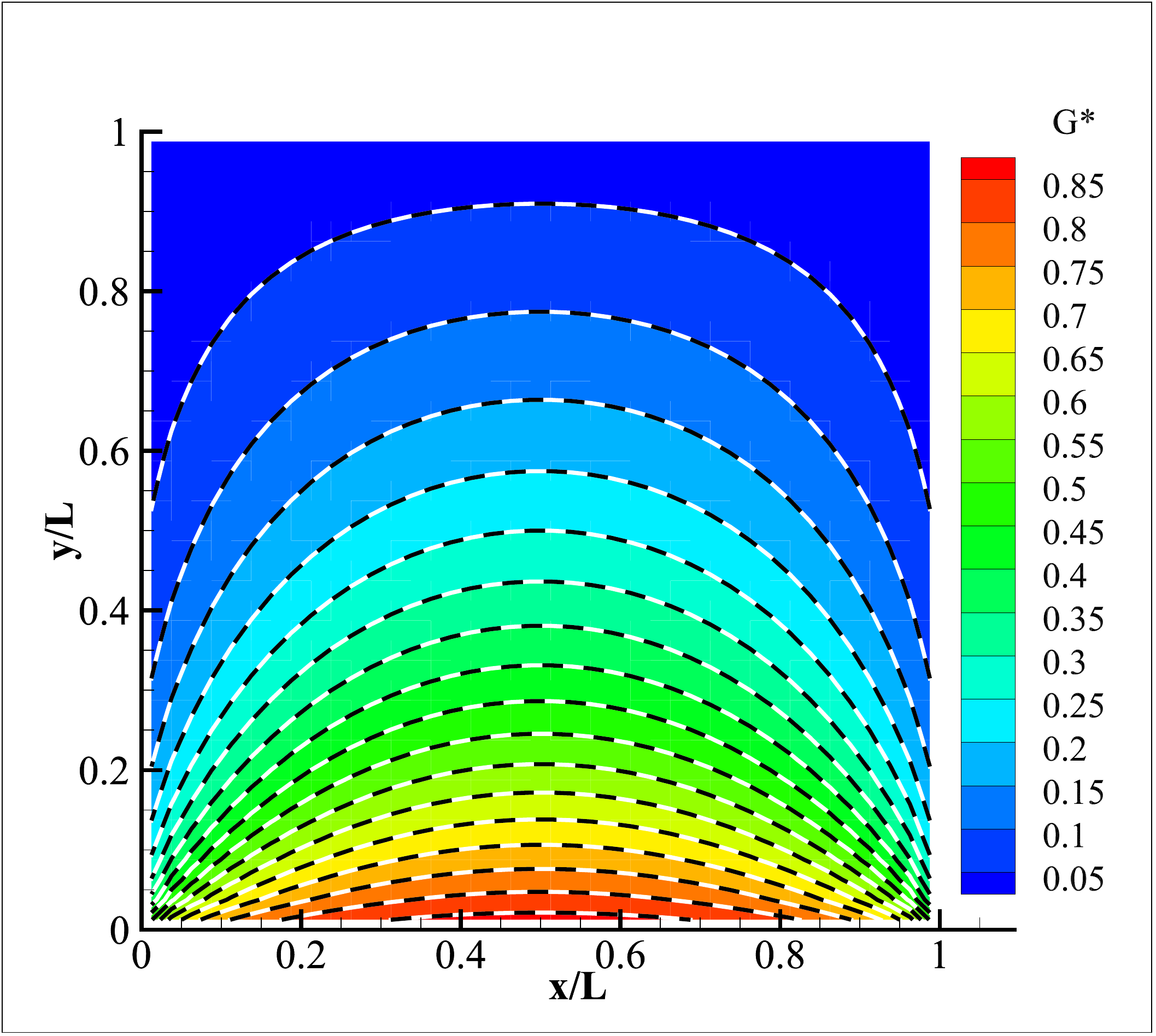}}~~
 \subfloat[]{\includegraphics[width=0.35\textwidth]{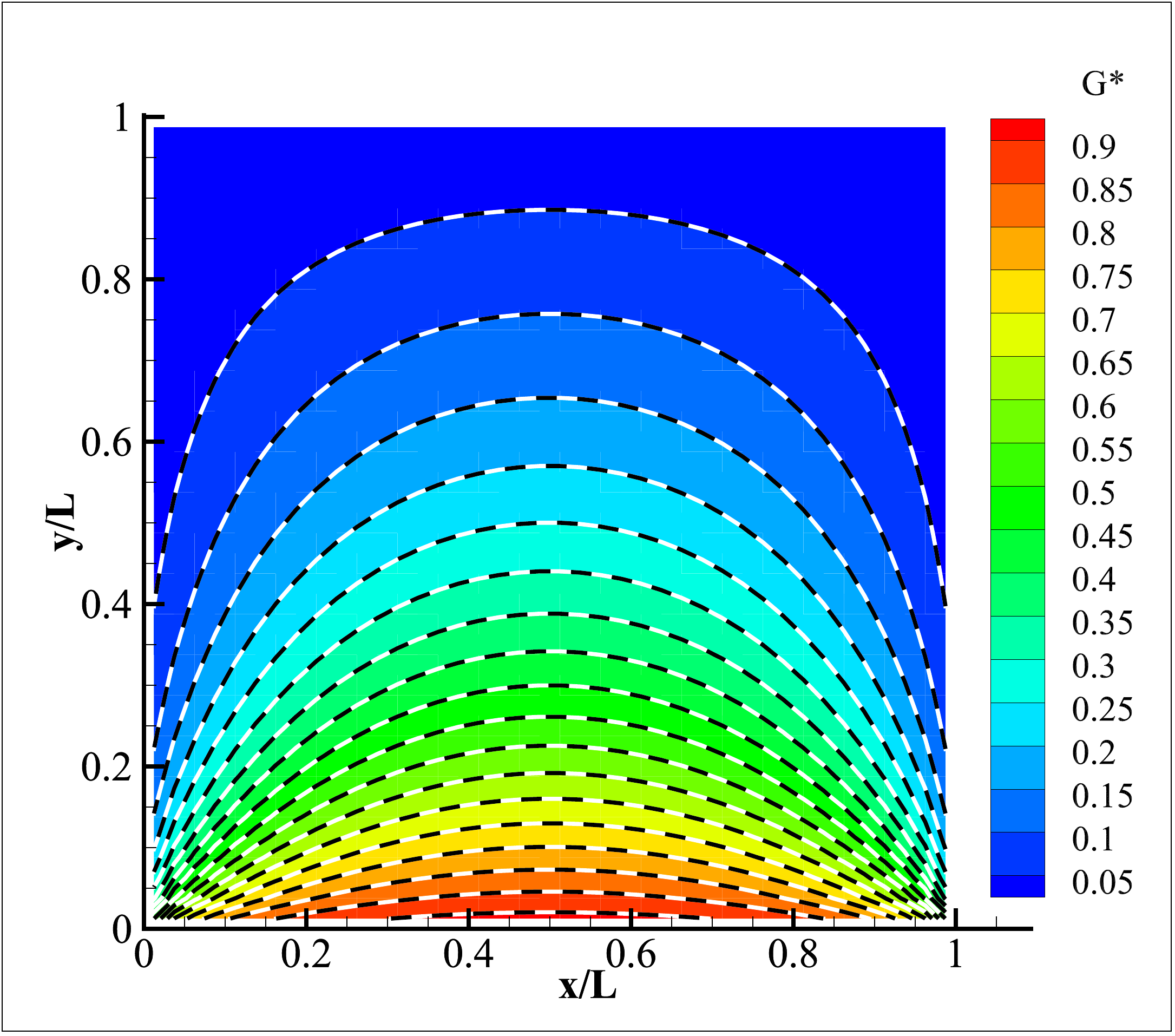}}~~
 \caption{Radiation energy distributions in the square enclosure with different optical thicknesses.(a) $\tau=1$, (b) $\tau=5$, (c) $\tau=10$ and (d) $\tau=20$. Background: DUGKS, dashed lines: T-SDUGKS.
 }
 \label{figure42}
\end{figure}
The scatter albedo is set to be $\omega =1$.
A uniform mesh with size $N_{x} \times N_{y}=40 \times 40$ is used. The direction cosine of zenith angle $\mu \in \left[ -1,1 \right]$ is discretized with the 16-point Gauss-Legendre quadrature, and the azimuth angle $\varphi \in \left[ 0,\pi \right]$ (due to geometric symmetry) is also discretized with the 16-point Gauss-Legendre quadrature.
The smooth linear interpolation is adopted when reconstructing the interface radiation intensity.

The radiation energy along the vertical centerline and heat flux at the bottom wall with different optical thicknesses are shown in Fig.~\ref{figure41}. It can be observed from Fig.~\ref{figure41}(b) that the results of the three DUGKS agree well with the reference results~\cite{luo2018multiscale}.
For further comparison, the radiation energy distributions from the DUGKS and T-SDUGKS are exhibited in Fig.~\ref{figure42}.
Clearly, the results predicted by the T-SDUGKS are in excellent agreement with those of the DUGKS.
\begin{table}[!htb]
\caption{Comparison of the computation costs of different schemes for the black square enclosure.}
\centering
\begin{tabular}{c c c c c c c c c c c}
  \hline
  \multirow{2}{*}{$\tau_L$} & \multicolumn{2}{c}{DUGKS}& \multicolumn{3}{c}{T-SDUGKS} & \multicolumn{3}{c}{R-SDUGKS}  \\
 \cline{2-9}
   &Time(s) &  Steps & Time(s)& Steps &Rate &  Time(s) &  Steps &Rate  \\
 \hline
 0.1	&28.066	  &2199	  &3.462	&178	&8.1	&2.775	  &178	  &10.1  \\
1		&14.499	  &1099	  &1.868	&97		&7.8	&1.641	  &97	  &8.8   \\
2		&15.132	  &1184	  &1.618	&79		&9.4	&1.314	  &79	  &11.5  \\
5		&22.632	  &1775	  &2.815	&151	&8.0	&2.382	  &151	  &9.5   \\
10		&35.353	  &2788	  &7.534	&416	&4.7	&6.121	  &416	  &5.8   \\
20		&60.252	  &4759	  &23.011	&1317	&2.6	&19.062	  &1317	  &3.2   \\
 \hline
\end{tabular}
\label{tab3}
\end{table}
The computation time and steps of different methods are listed in Table~\ref{tab3}. It can be seen that the T-SDUGKS and R-SDUGKS have faster convergence rates than the DUGKS. The comparisons show that the T-SDUGKS and R-SDUGKS are accurate and more efficient than DUGKS for steady radiative heat transfer problems.
\subsection{Radiation in a black square enclosure with different media}\label{sec45}

\begin{figure}[!htb]
    \centering
    \includegraphics[width=0.4\textwidth]{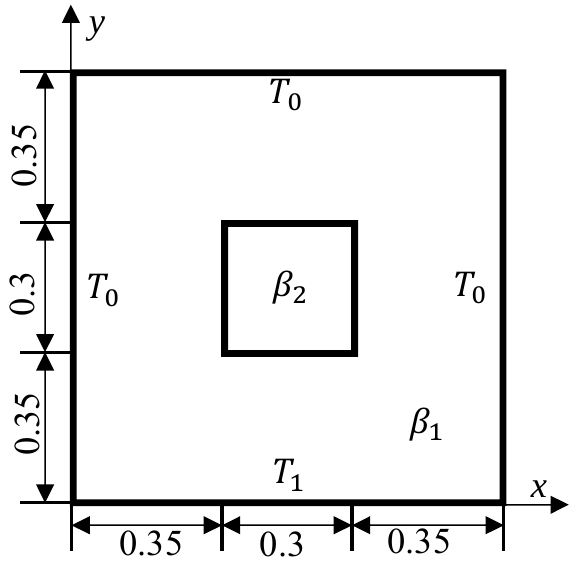}
    \centering
    \caption{Schematic of two-dimensional radiation heat transfer.}
\label{tufour}
\end{figure}

The radiation in a black square enclosure filled with two different media is considered in this section.
This case is used to show that the slope limiter for reconstructing the interface radiation intensity is preferred for problems with discontinuous extinction coefficients.
As shown in Fig.~\ref{tufour}, a black square with the side length of $L=1$ is filled with two different media.
The nondimensional emissive power is set to be $\Phi_{1}=1 $ at the bottom wall, and $\Phi_{0}=0 $ at the other walls and the media.
The extinction coefficients of the inner and outer media are $\beta_{2}$ and $\beta_{1}$, respectively.
In the simulations, the scatter albedo is set to be $\omega =1$. $\beta_{1}$  ranges from 0.01 to 100 and $\beta_{2}=1/\beta_{1}$.
\begin{figure}[!hb]
 \centering
\subfloat[]{\includegraphics[width=0.35\textwidth]{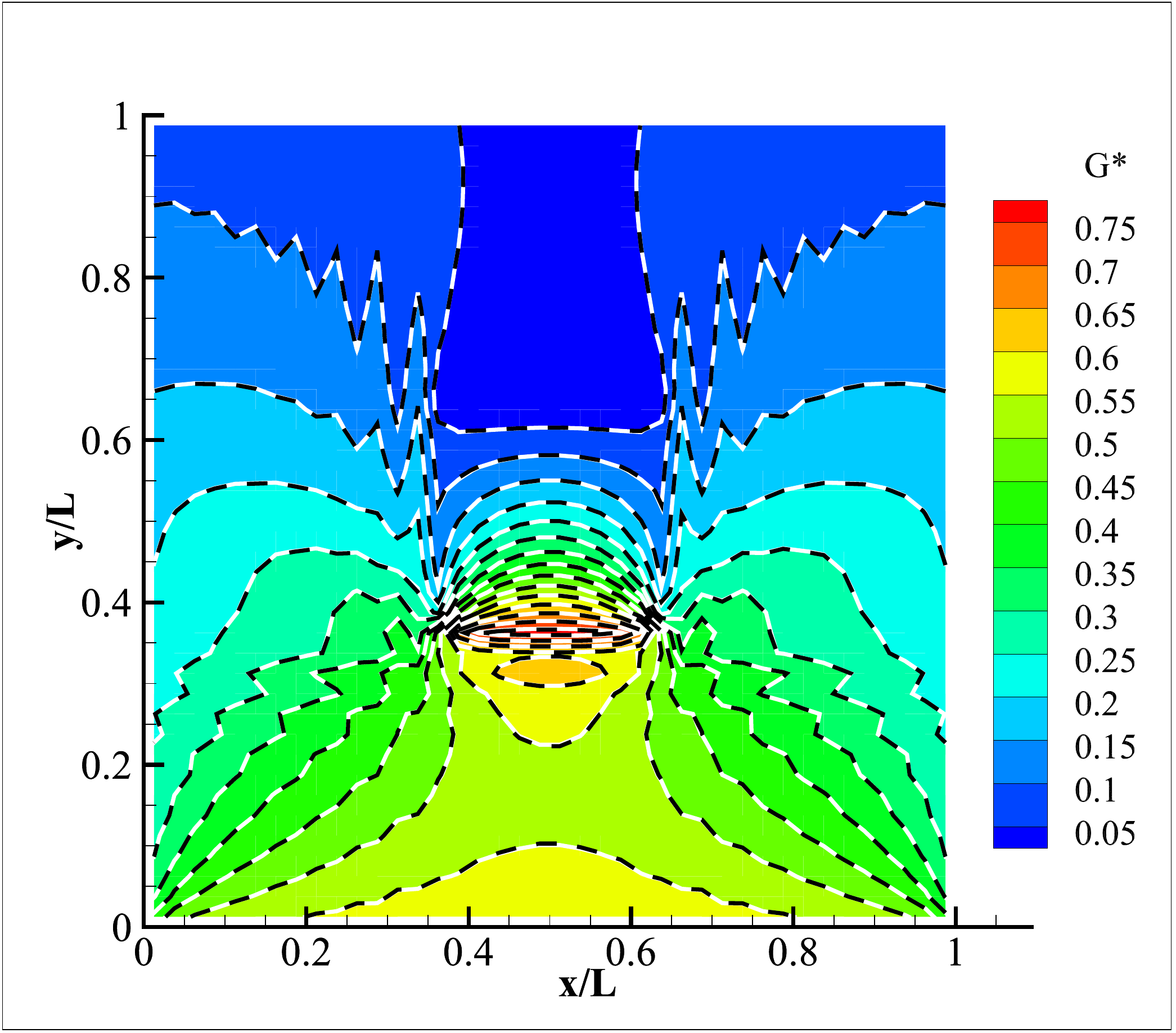}}~~
 \subfloat[]{\includegraphics[width=0.35\textwidth]{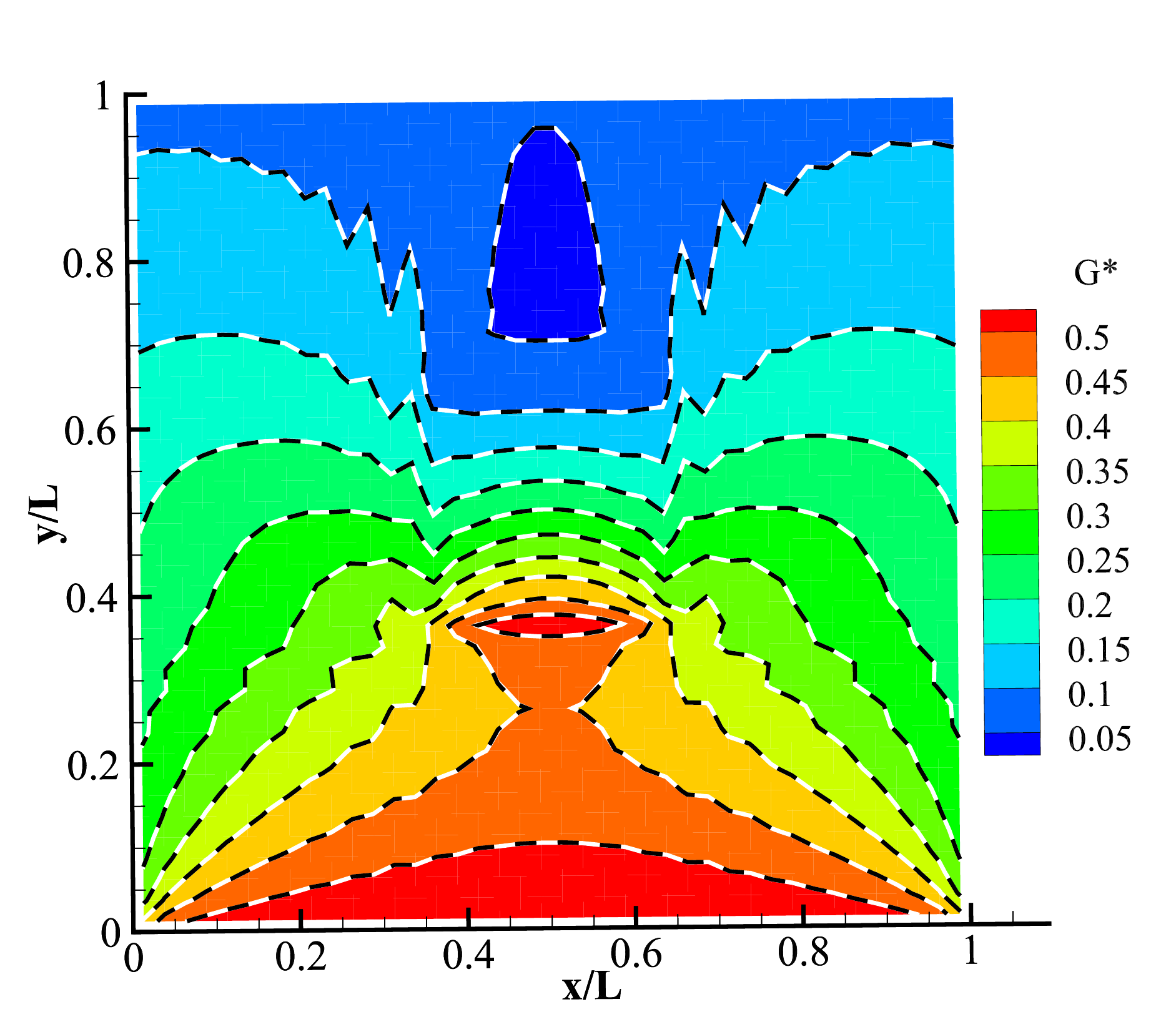}}~~\\
 \subfloat[]{\includegraphics[width=0.35\textwidth]{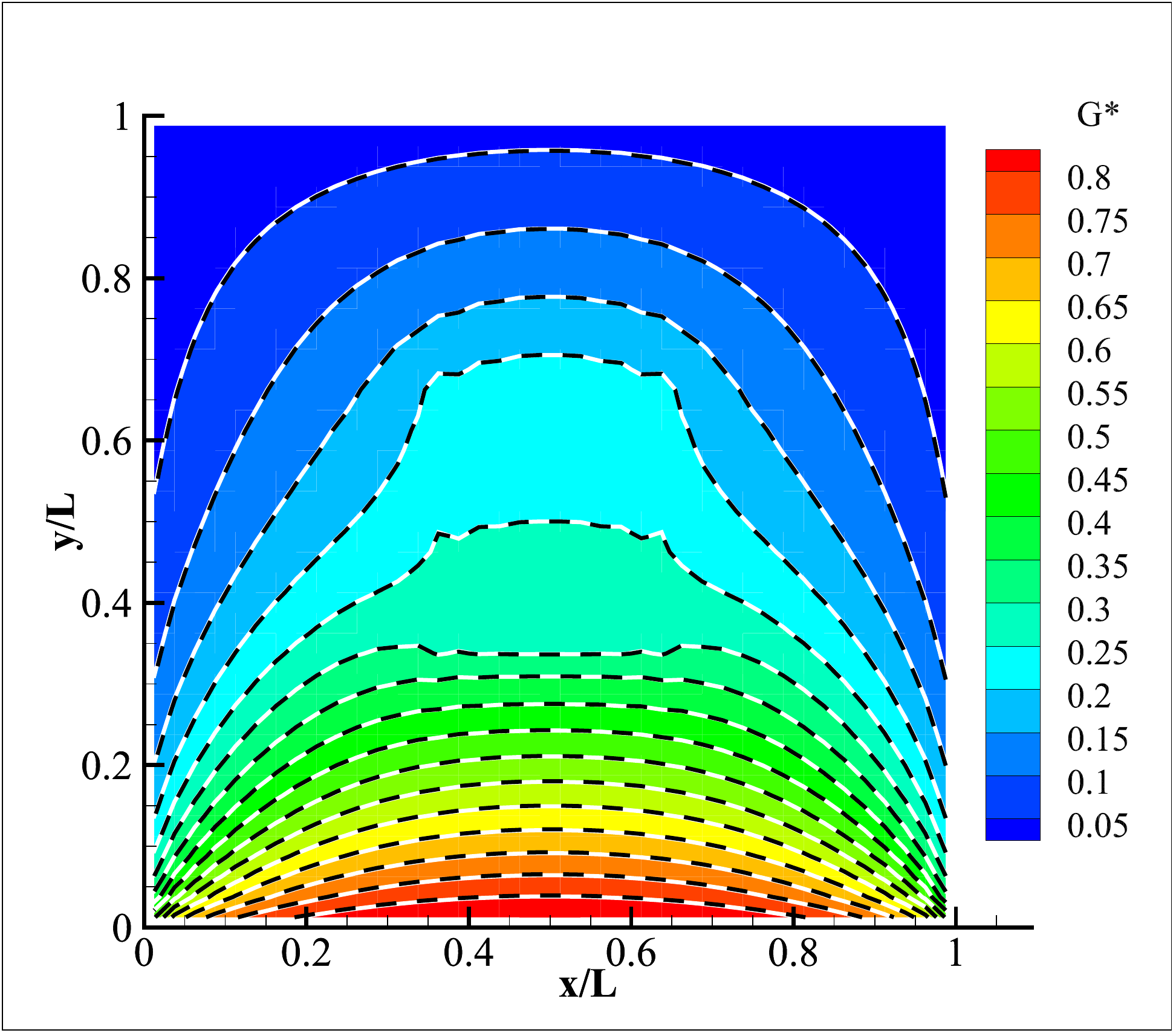}}~~
 \subfloat[]{\includegraphics[width=0.35\textwidth]{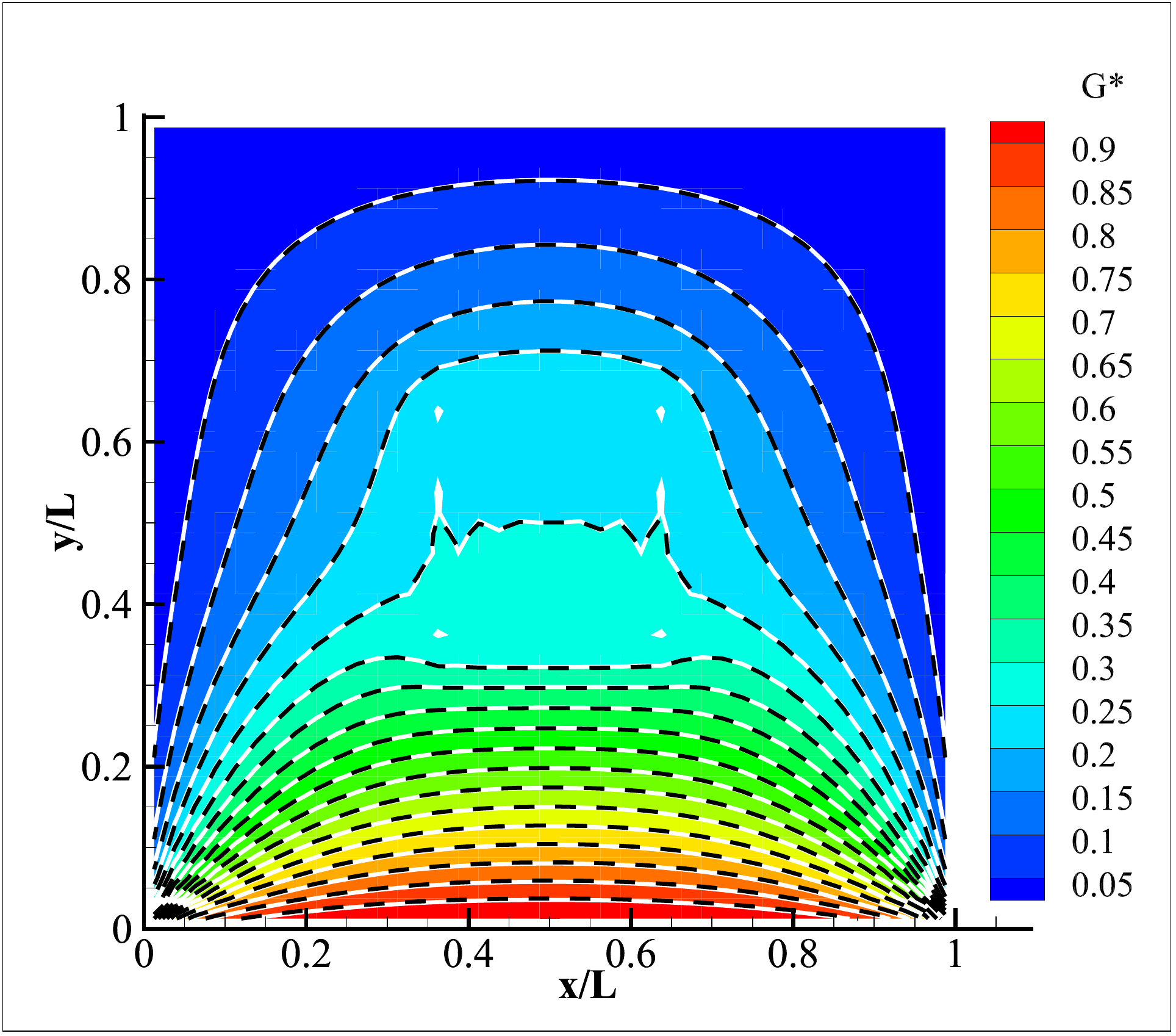}}~~
 \caption{  Radiation energy distributions in the square enclosure with smooth linear interpolation. (a) $\beta_{1}=0.01,\beta_{2}=100$, (b)  $\beta_{1}=0.1,\beta_{2}=10$, (c) $\beta_{1}=10,\beta_{2}=0.1$ and (d) $\beta_{1}=100,\beta_{2}=0.01$. Background: T-SDUGKS, dashed lines: R-SDUGKS.
 }
 \label{figure54}
\end{figure}
In order to reduce the ray effect, the solid angle is discretized using the Gauss-Legendre quadrature with a $N_{\mu} \times N_{\varphi}=32 \times 32$ angular mesh.

We first test the T-SDUGKS using the smooth linear reconstruction. The radiation energy with a mesh $N_{x} \times N_{y}=40 \times 40$ is displayed in Fig.~\ref{figure54}. Clearly, numerical oscillations appear in the optical thin area, indicating that the smooth linear interpolation is not suitable for problems with discontinuous extinction coefficients.

\begin{figure}[!hb]
 \centering
\subfloat[]{\includegraphics[width=0.35\textwidth]{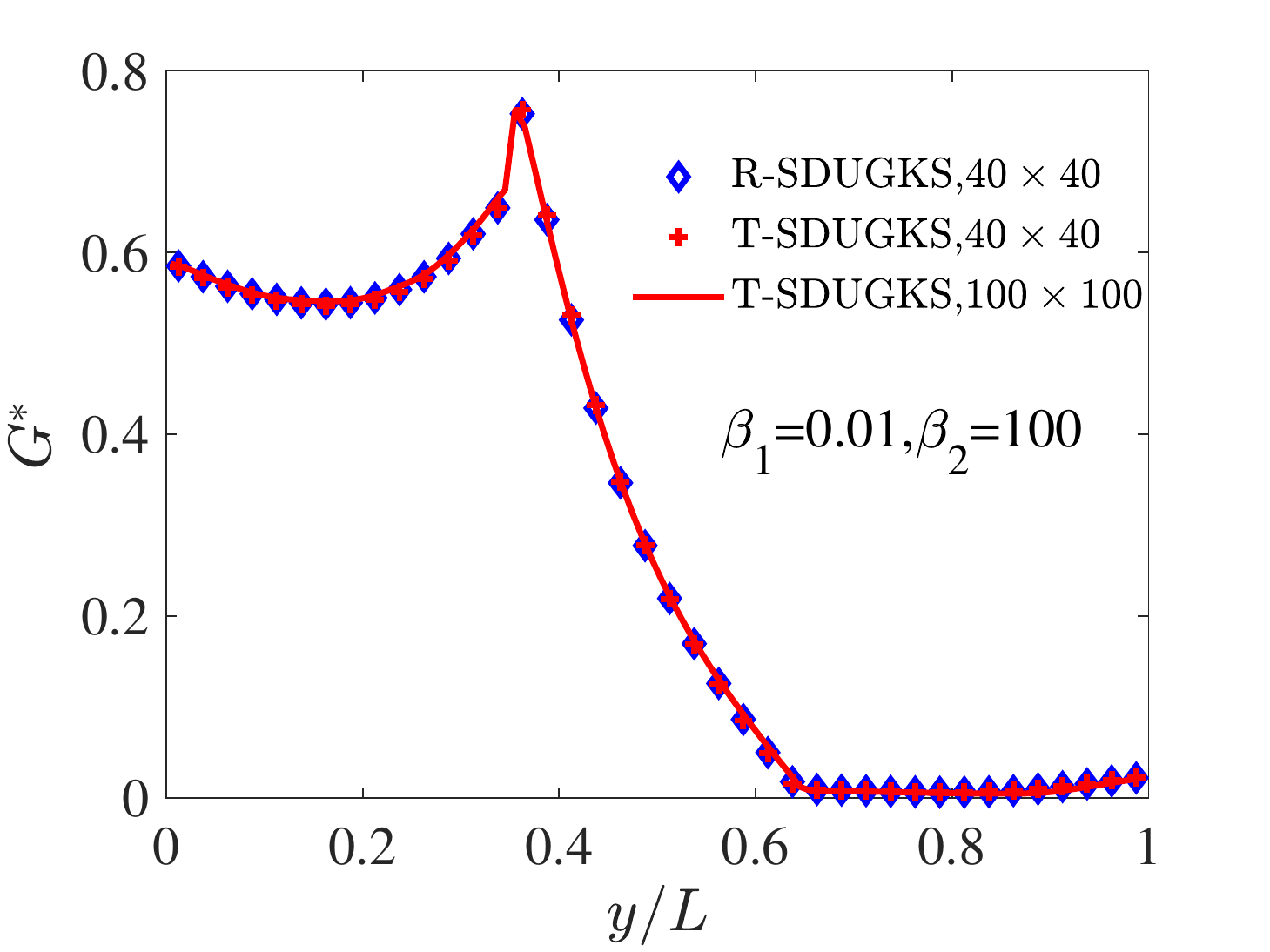}}~~
 \subfloat[]{\includegraphics[width=0.35\textwidth]{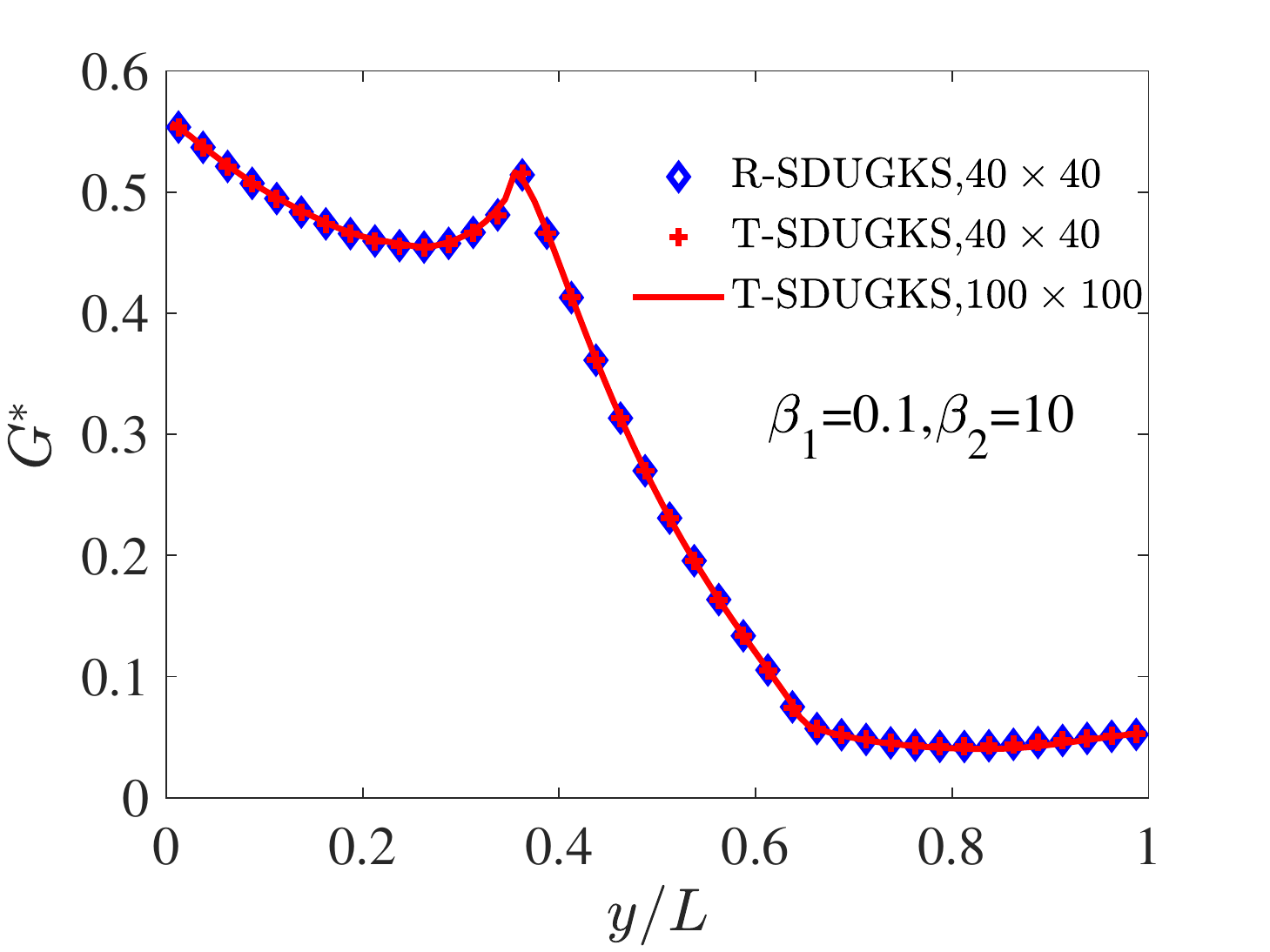}}~~\\
 \subfloat[]{\includegraphics[width=0.35\textwidth]{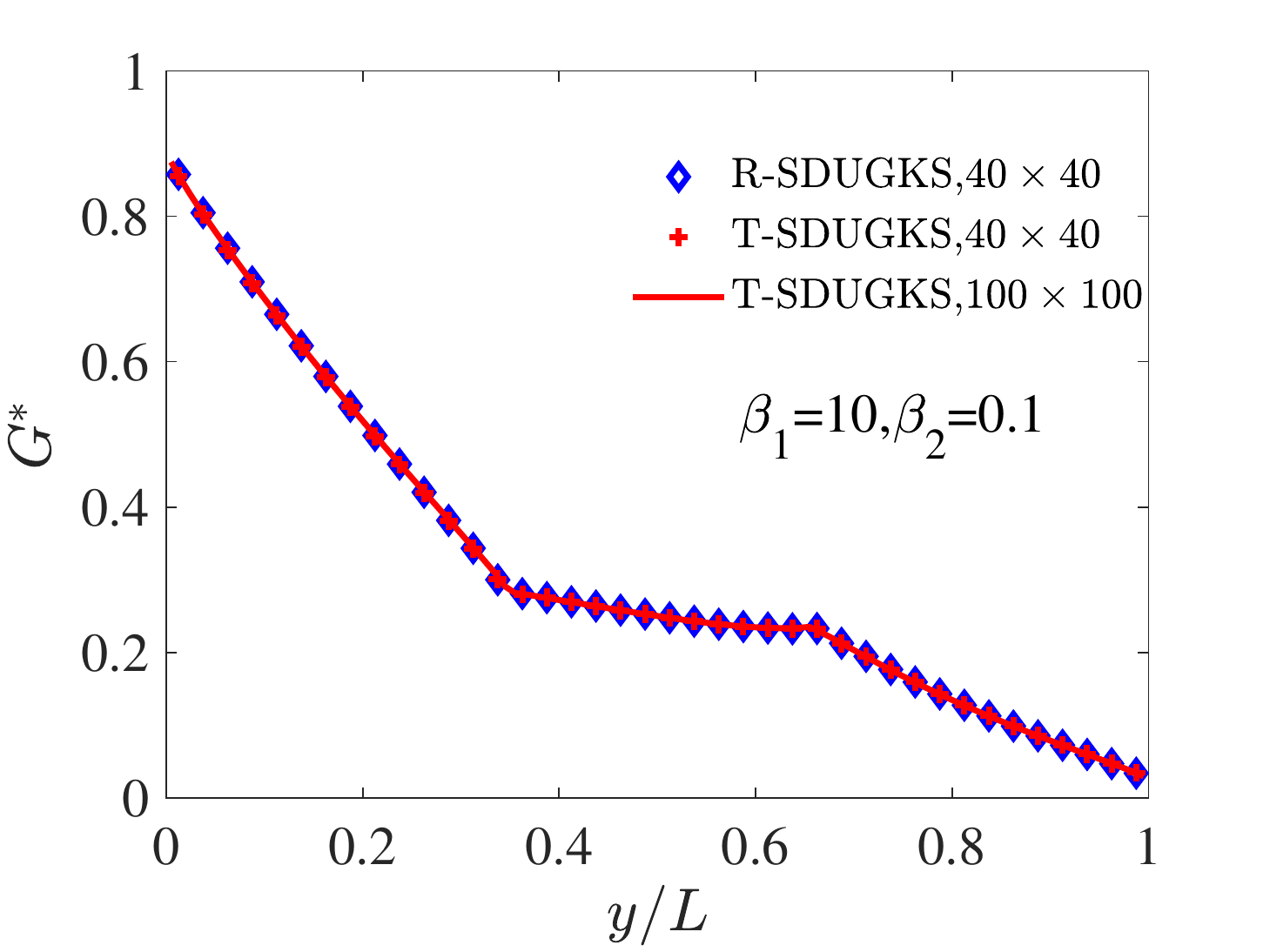}}~~
 \subfloat[]{\includegraphics[width=0.35\textwidth]{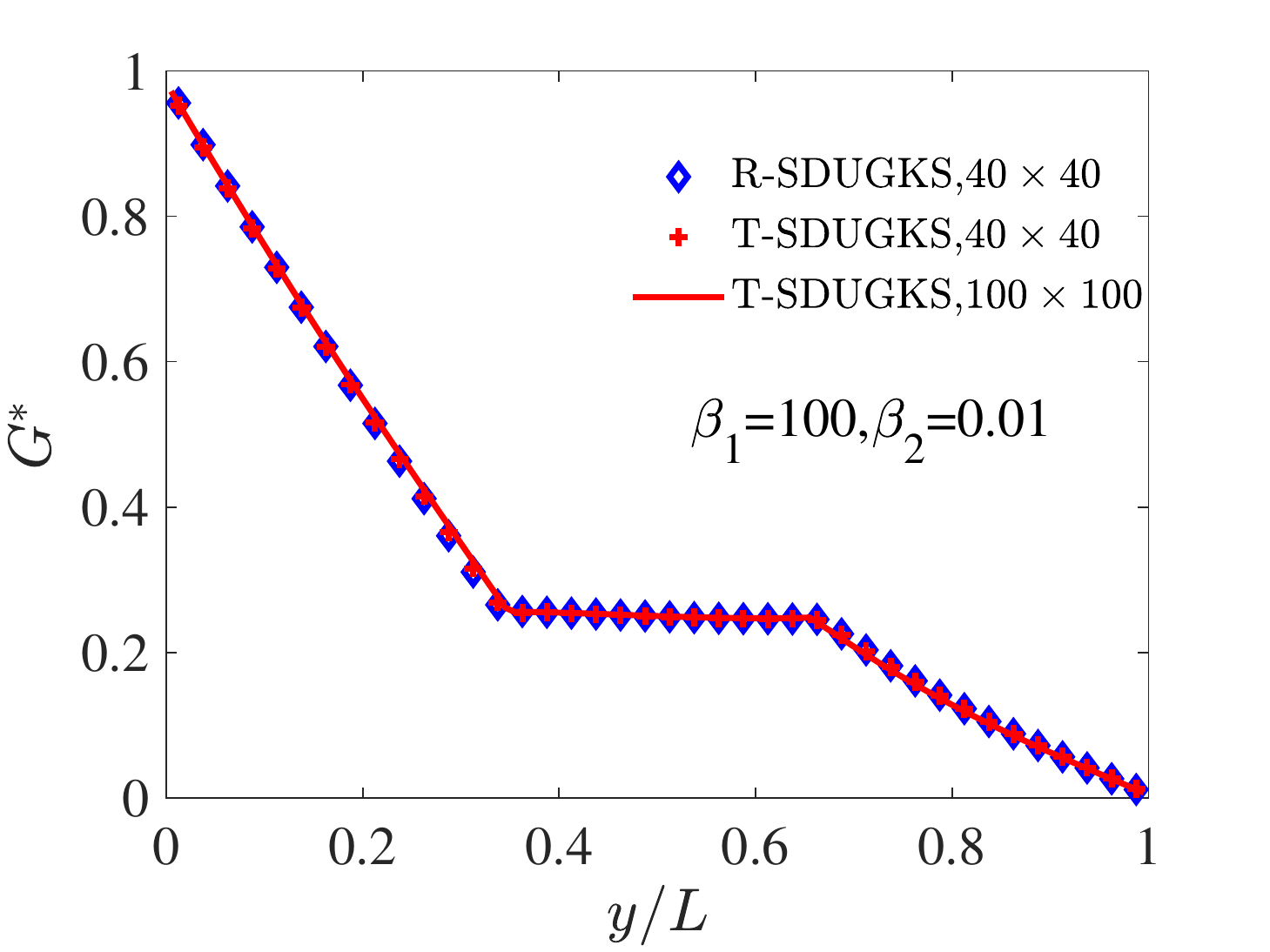}}~~
 \caption{Radiation energy profiles along the vertical centerline with different optical thicknesses and grid sizes.
 }
 \label{figure51}
\end{figure}

\begin{figure}[!htb]
 \centering
\subfloat[]{\includegraphics[width=0.35\textwidth]{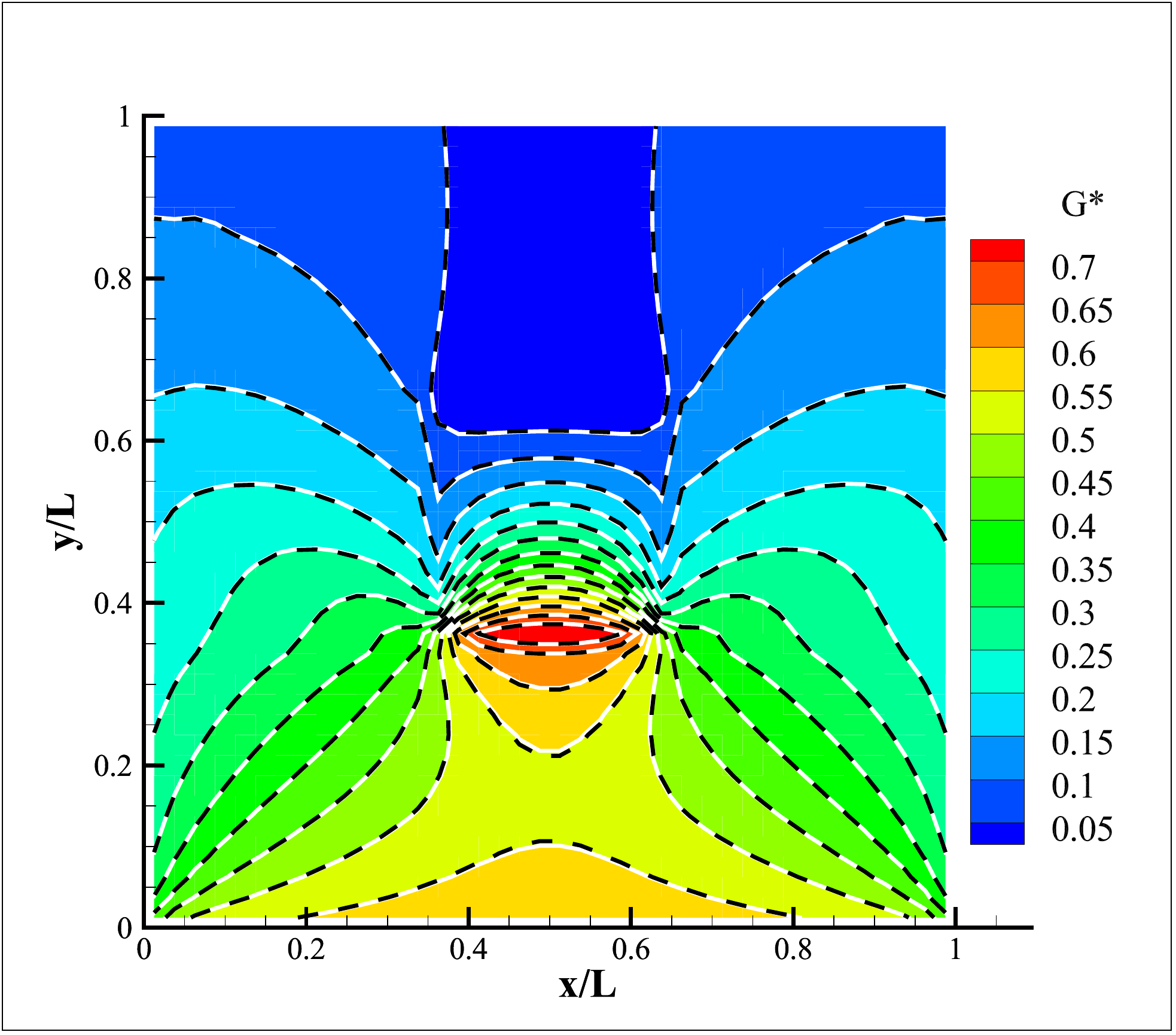}}~~
 \subfloat[]{\includegraphics[width=0.35\textwidth]{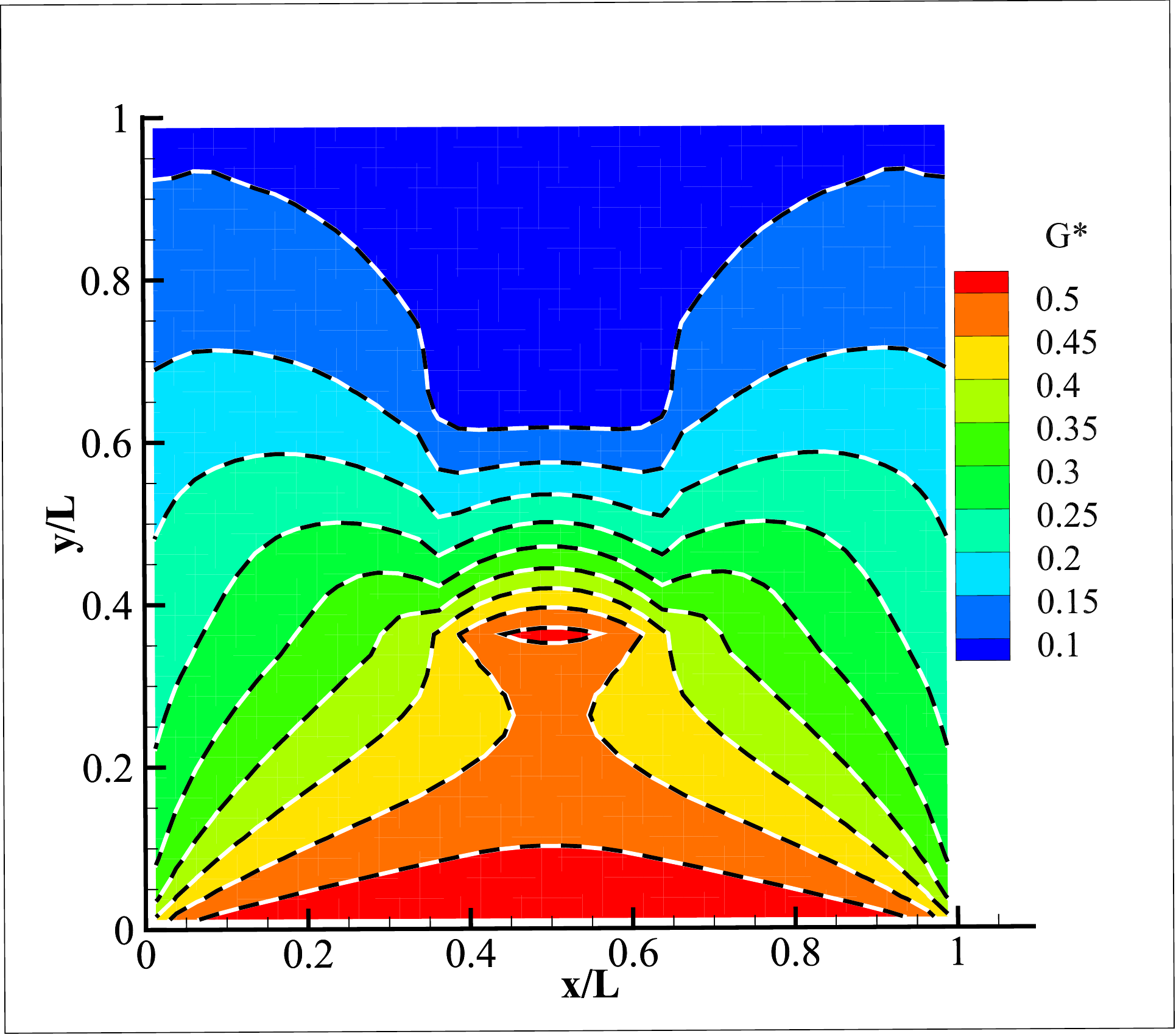}}~~\\
 \subfloat[]{\includegraphics[width=0.35\textwidth]{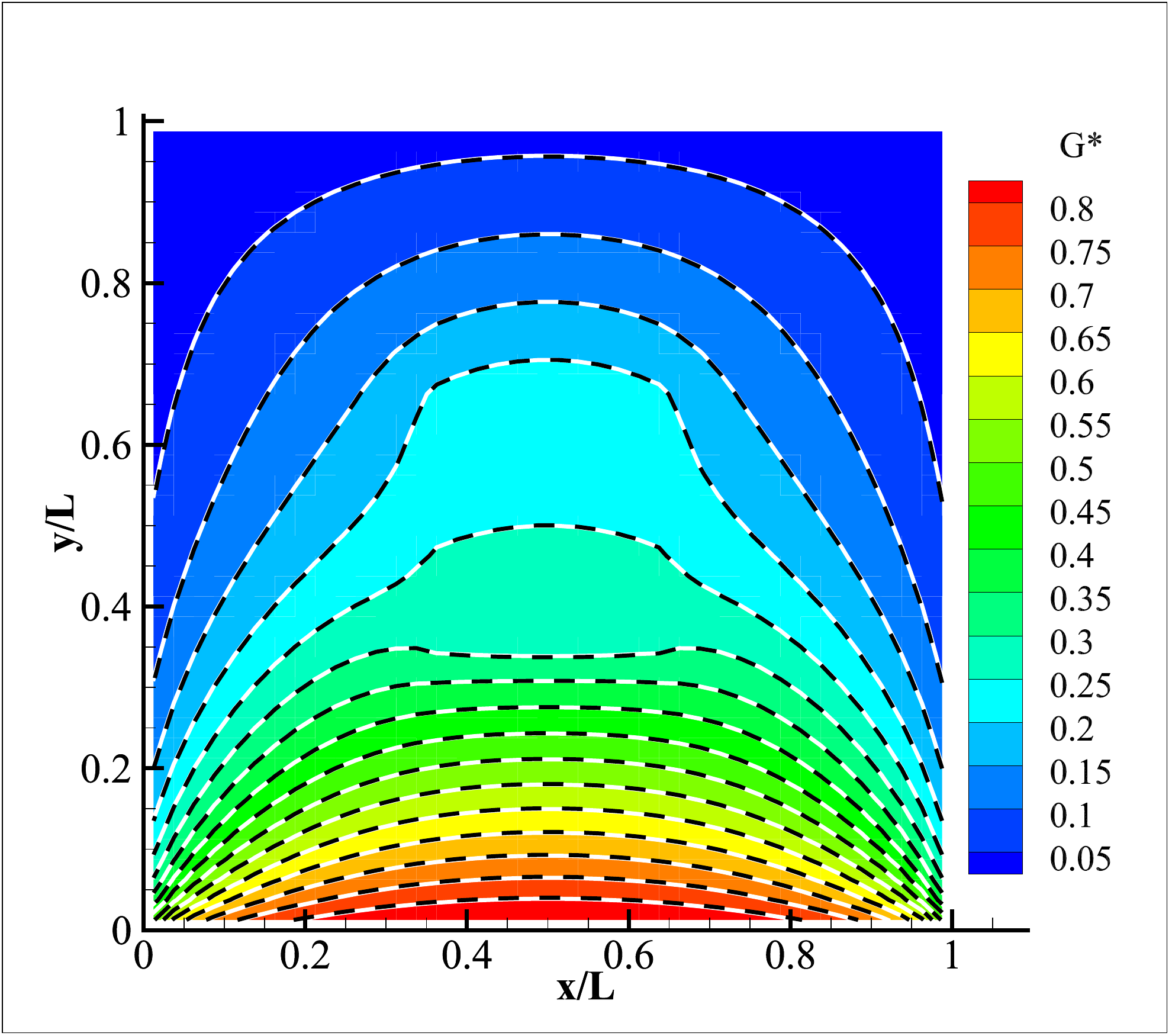}}~~
 \subfloat[]{\includegraphics[width=0.35\textwidth]{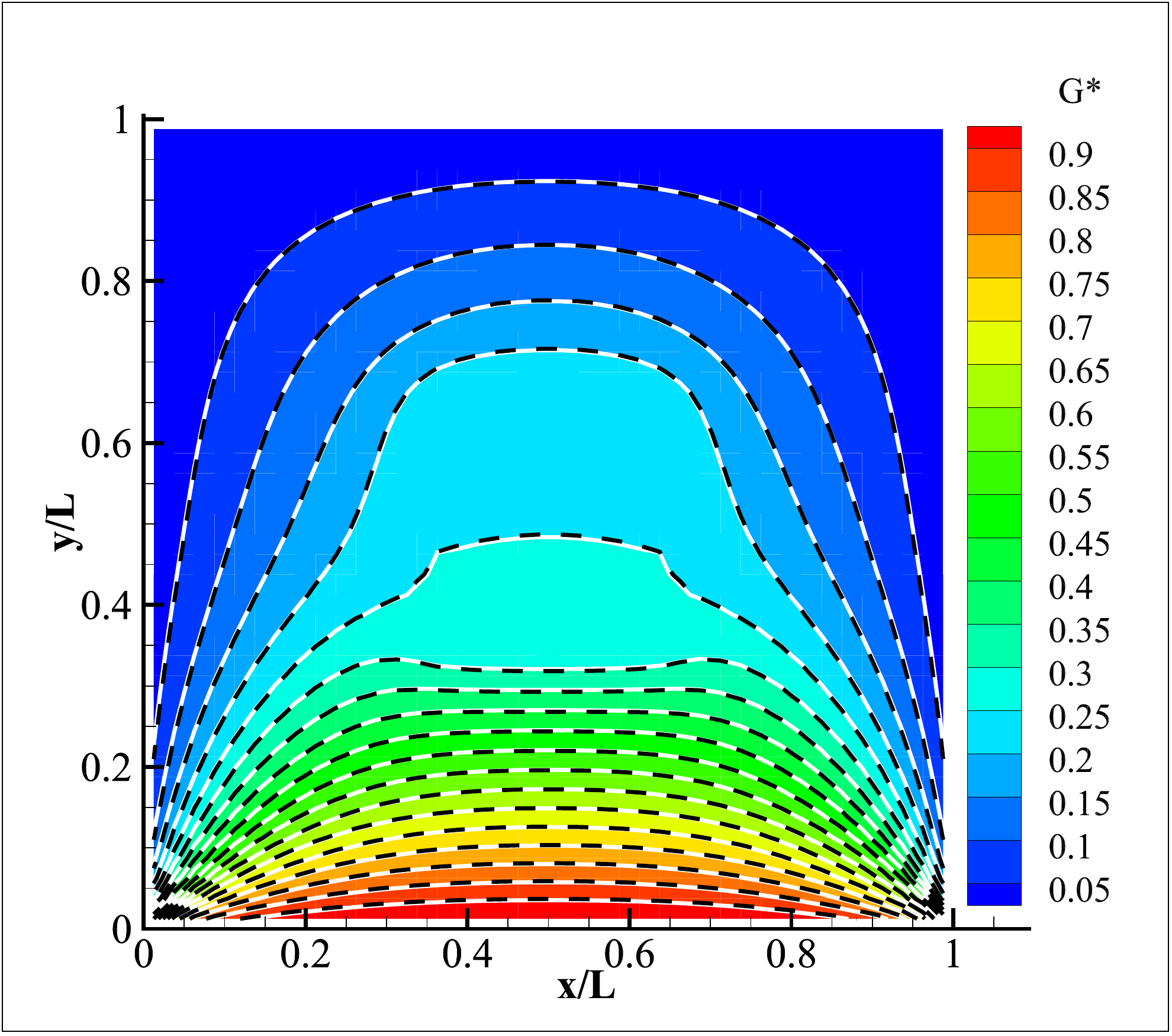}}~~
 \caption{Radiation energy distributions in the square enclosure with van Leer limiter. (a) $\beta_{1}=0.01,\beta_{2}=100$, (b)  $\beta_{1}=0.1,\beta_{2}=10$, (c) $\beta_{1}=10,\beta_{2}=0.1$ and (d) $\beta_{1}=100,\beta_{2}=0.01$. Background: T-SDUGKS, dashed lines: R-SDUGKS.
 }
 \label{figure53}
\end{figure}

The T-SDUGKS using linear reconstruction with van Leer limiter is also applied to this problem. Figure~\ref{figure51} shows the radiation energy along the vertical centerline with different optical thicknesses and grid sizes, and Fig.~\ref{figure53} shows the contours of the radiation energy distributions. It can be seen that numerical oscillations disappear.
From Fig.~\ref{figure53}, it can be seen that the extinction coefficient of the inner medium has a great influence on the transport of radiation energy. When $\beta_{2} > \beta_{1}$, a hot zone appears between the inner and outer media close to the hot wall. This can be attributed to the strong photon scattering effect of the inner optical thick medium.
As $\beta_{2} < \beta_{1}$, the thickness of the inner medium is smaller and photon scatterings are rare, leading to weak energy exchange.
The above results demonstrate that the T-SDUGKS using the van Leer reconstruction is an accurate method for steady multiscale radiative heat transfer problems, while the smooth reconstruction is inappropriate for problems with discontinuous extinction coefficients.
\subsection{Radiation in a slab with a heat source}\label{sec43}
In this section the performance of the T-SDUGKS and the R-SDUGKS is further compared by simulating the radiation transfer in a slab of size $L=1$ with a constant heat source, which was first considered in Ref.~\cite{mieussens2013asymptotic}.
The transfer equation for this problem reads,
\begin{equation}
   \mu_{k}  \cdot \frac{dI\left(x , \mu_{k} \right)}{dx} = -\beta \left(  x \right)  I \left( x , \mu_{k} \right)+ \frac{\beta \left(  x \right)}{2} \sum_{m=1}^{M} I \left(x, \mu_m \right) \omega_m+\frac{\dot{Q}}{2}.
\label{eq:RTEGeneration}
\end{equation}
The left and right walls of the slab are kept cold with $\Phi_{0}=0$ and the heat source $\dot{Q}=0.01$.
A 40-point Gauss-Legendre quadrature is used to discretize the solid angle space, and uniform meshes with size $N_x =40,100,200$  are used for the physical space, respectively.
The reference solution is obtained by the DUGKS with a uniform mesh with 4000 cells.
\begin{figure}[!ht]
 \centering
\subfloat[]{\includegraphics[width=0.35\textwidth]{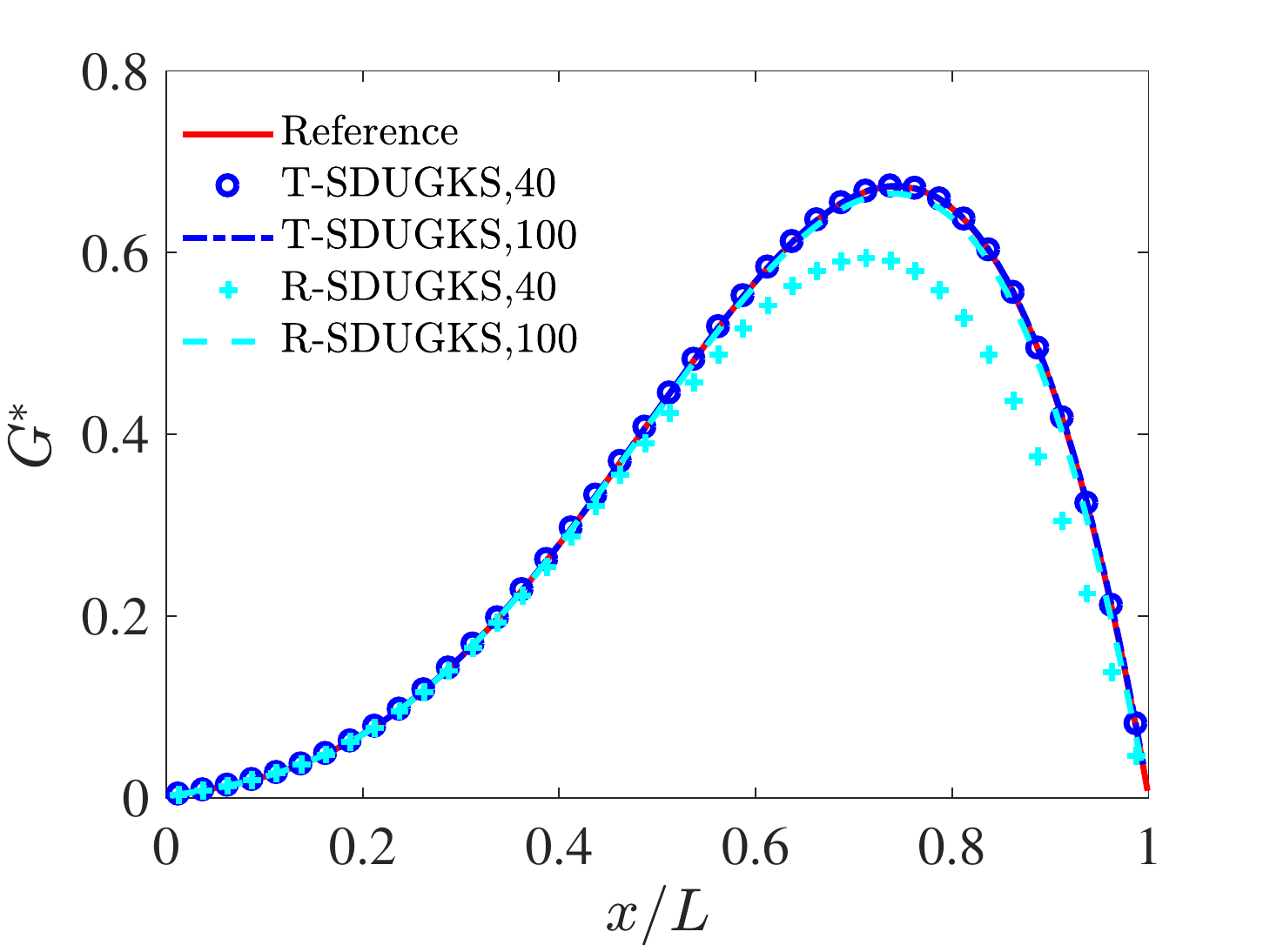}}~~
 \subfloat[]{\includegraphics[width=0.35\textwidth]{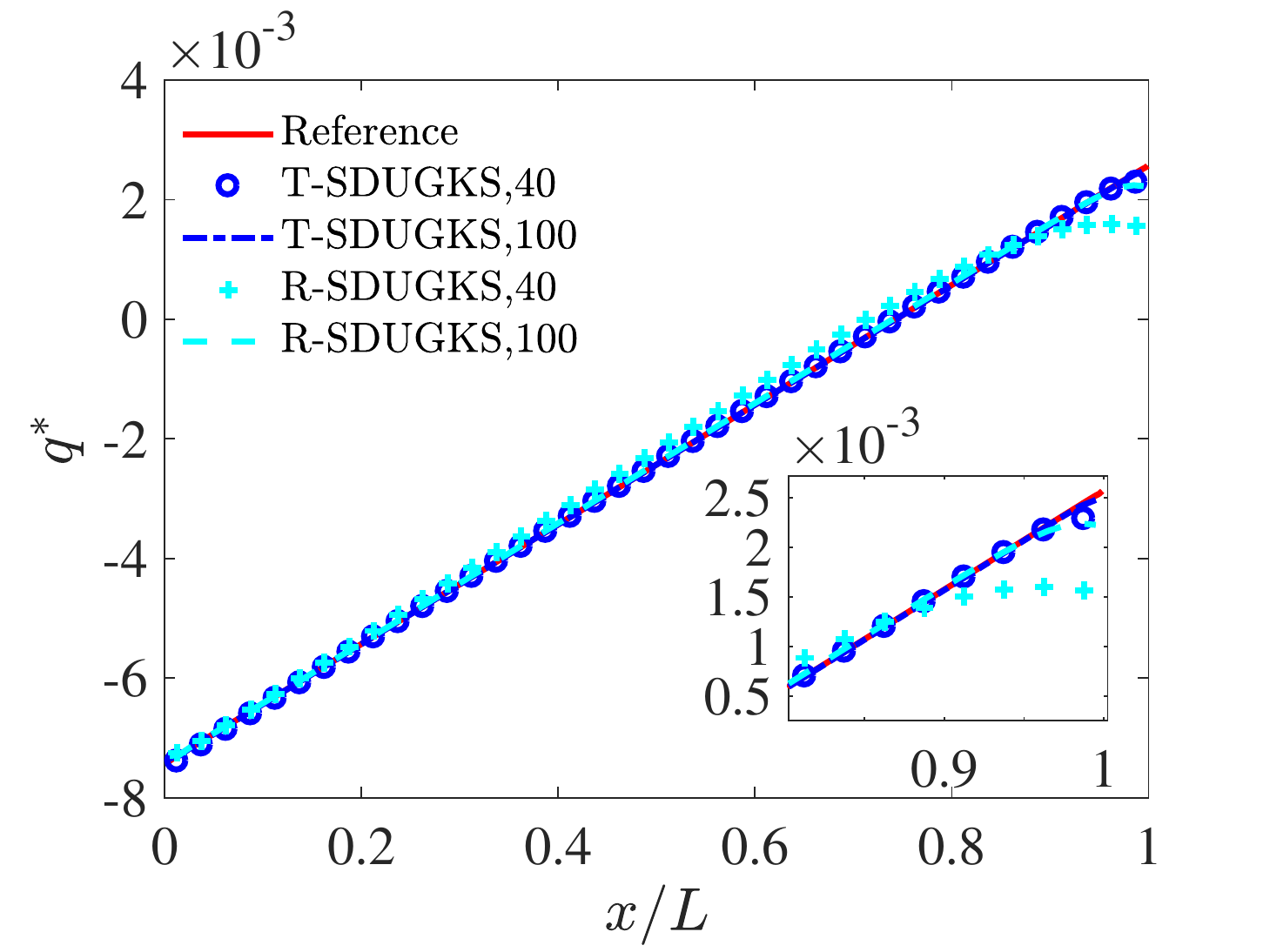}}~~
 \caption{Radiation energy and heat flux profiles of the T-SDUGKS and R-SDUGKS with $l=0.25dx$.
 }
 \label{figure31}
\end{figure}%

First, we simulate the case with $\beta \left( x \right)=100 \left( 1+ 100x^2 \right)$. The smooth linear interpolation is adopted when reconstructing the interface radiation intensity.
Figure~\ref{figure31} shows the radiation energy and heat flux with different meshes at $l=0.25 \Delta x$.
\begin{figure}[!htb]
 \centering
\subfloat[]{\includegraphics[width=0.35\textwidth]{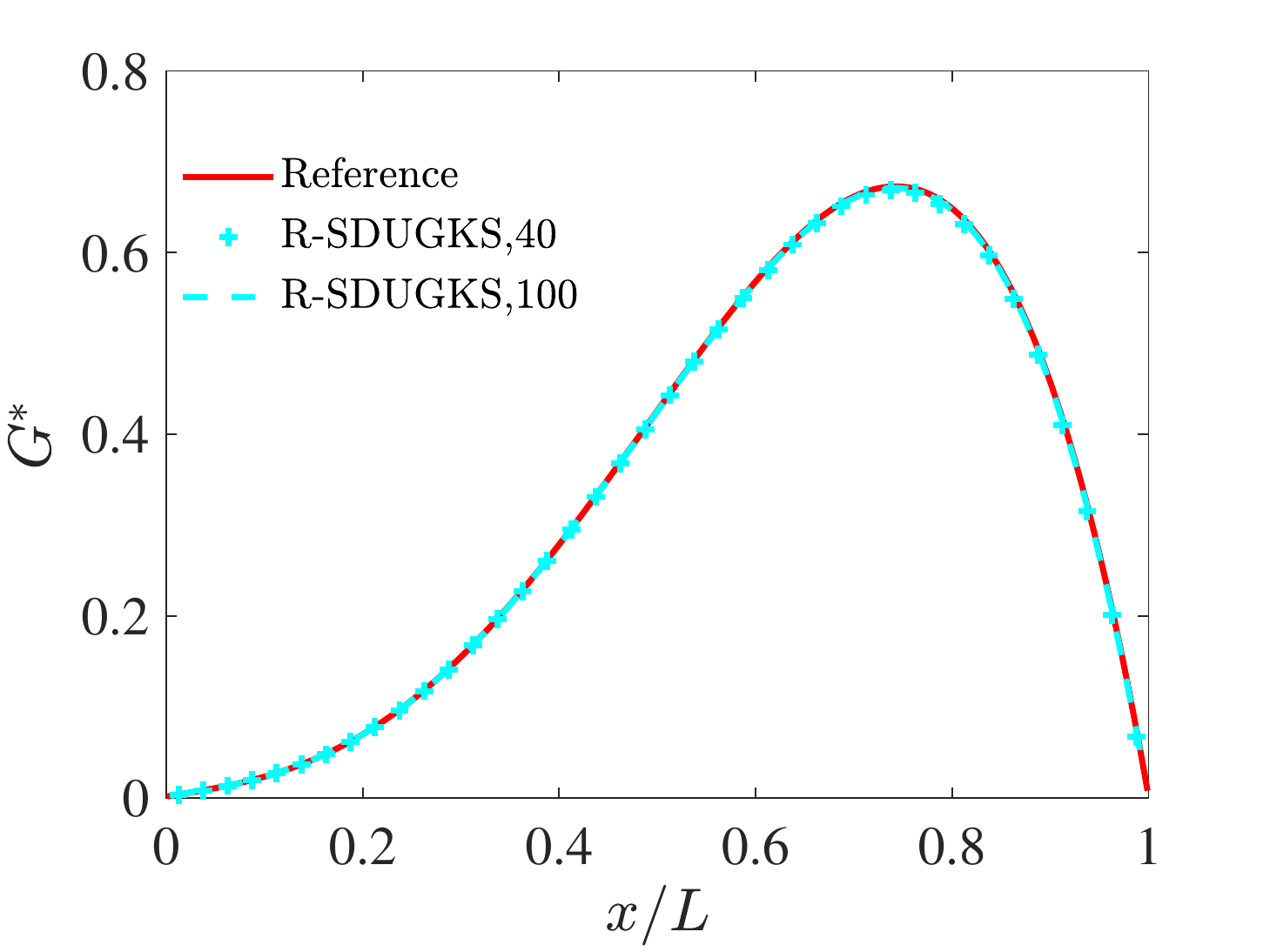}}~~
 \subfloat[]{\includegraphics[width=0.35\textwidth]{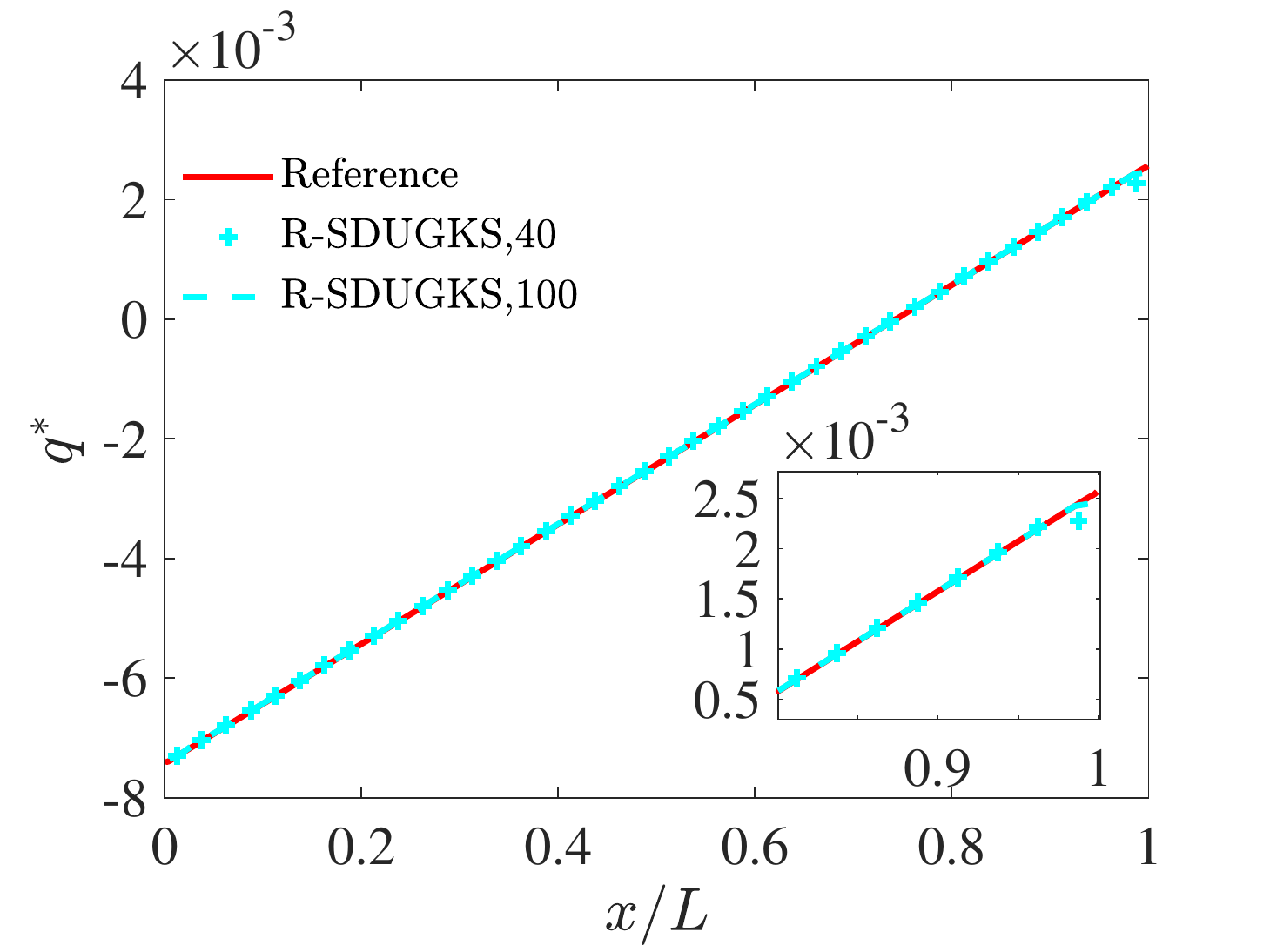}}~~
 \caption{Radiation energy and heat flux profiles of the R-SDUGKS with $l=1/\beta_{max}$.
 }
 \label{figure32}
\end{figure}%
Good agreement between the T-SDUGKS and the reference results can be observed, while significant deviations between the results of the R-SDUGKS and reference data can be found with the coarse mesh ($N_x=40$) for  radiation energy and heat flux.
The deviations in the R-SDUGKS may be caused by the fact that the characteristic line length is larger than $1/\beta_{max}$~\cite{zhou2020discrete}.
Therefore, $l =1/\beta_{max}$ is set for R-SDUGKS, which leads to a much smaller approximation error $O \left( {l^2} \right)$ in Eq.~\eqref{eq:rewrRectanFace}.
The corresponding results are shown in Fig.~\ref{figure32}. It is seen that the results agree well with the reference solutions, indicating that the characteristic line length in the rectangular SDUGKS is  limited by the extinction coefficient, while the T-SDUGKS dose not suffer from this limitation.

Furthermore, the case with a discontinuous extinction coefficient is also considered, where
\begin{equation}
   \beta \left( x \right)=
    \left\{
    \begin{array}{rcl}
    100,      & {0 \le x \le 0.1}, \\
    1000,     & {0.1 < x \le 0.5}, \\
    10000,    & {0.5 < x \le 1.0}.
    \end{array} \right.
\label{eq:DisconExtin}
\end{equation}
The characteristic line length in T-SDUGKS is chosen to be $l=0.25 \Delta x$, while in R-SDUGKS it is chosen as $l=1/\beta_{max}$.
The van Leer limiter is adopted for this discontinuous problem.
\begin{figure}[!htb]
 \centering
\subfloat[]{\includegraphics[width=0.35\textwidth]{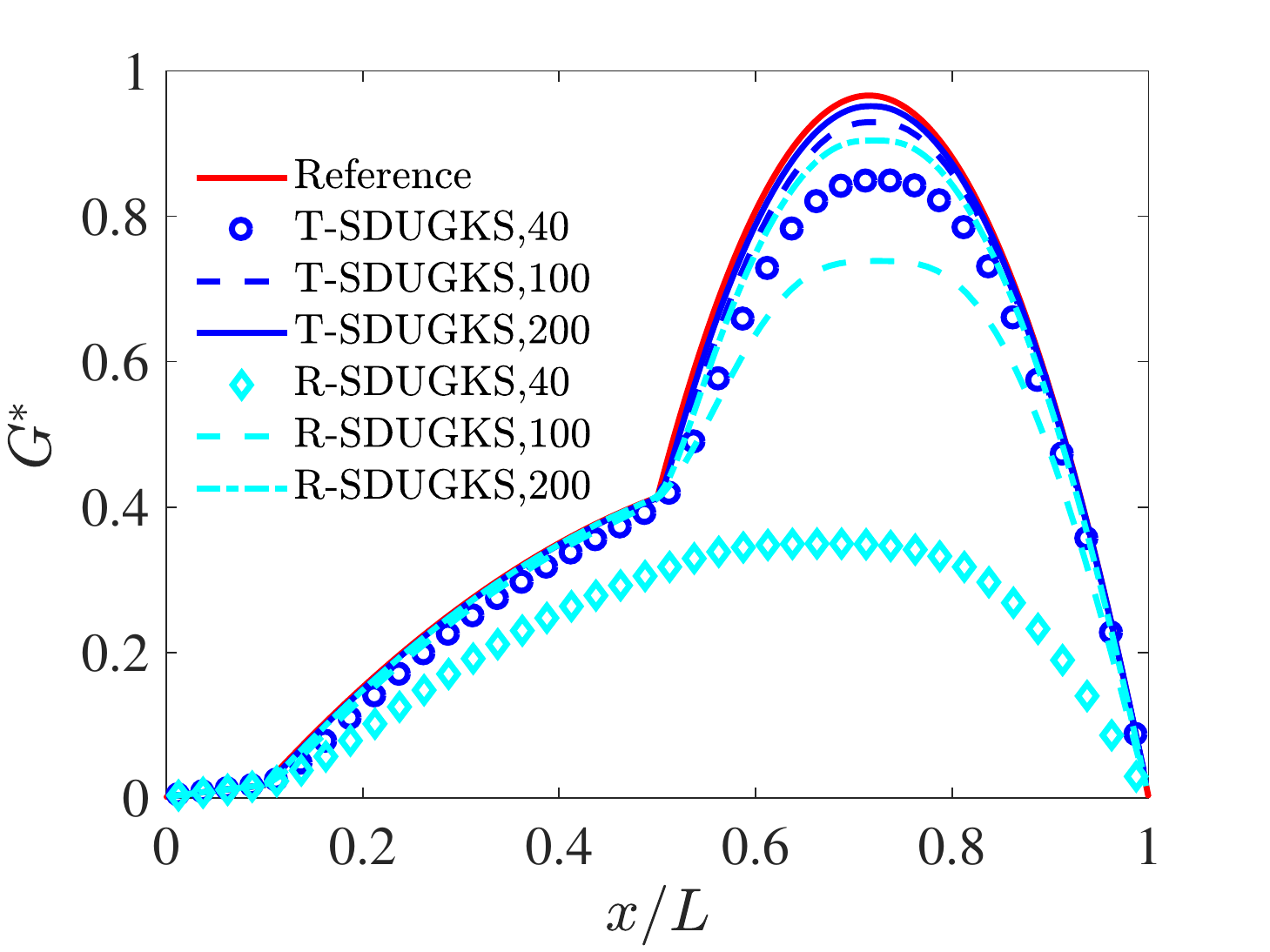}}~~
 \subfloat[]{\includegraphics[width=0.35\textwidth]{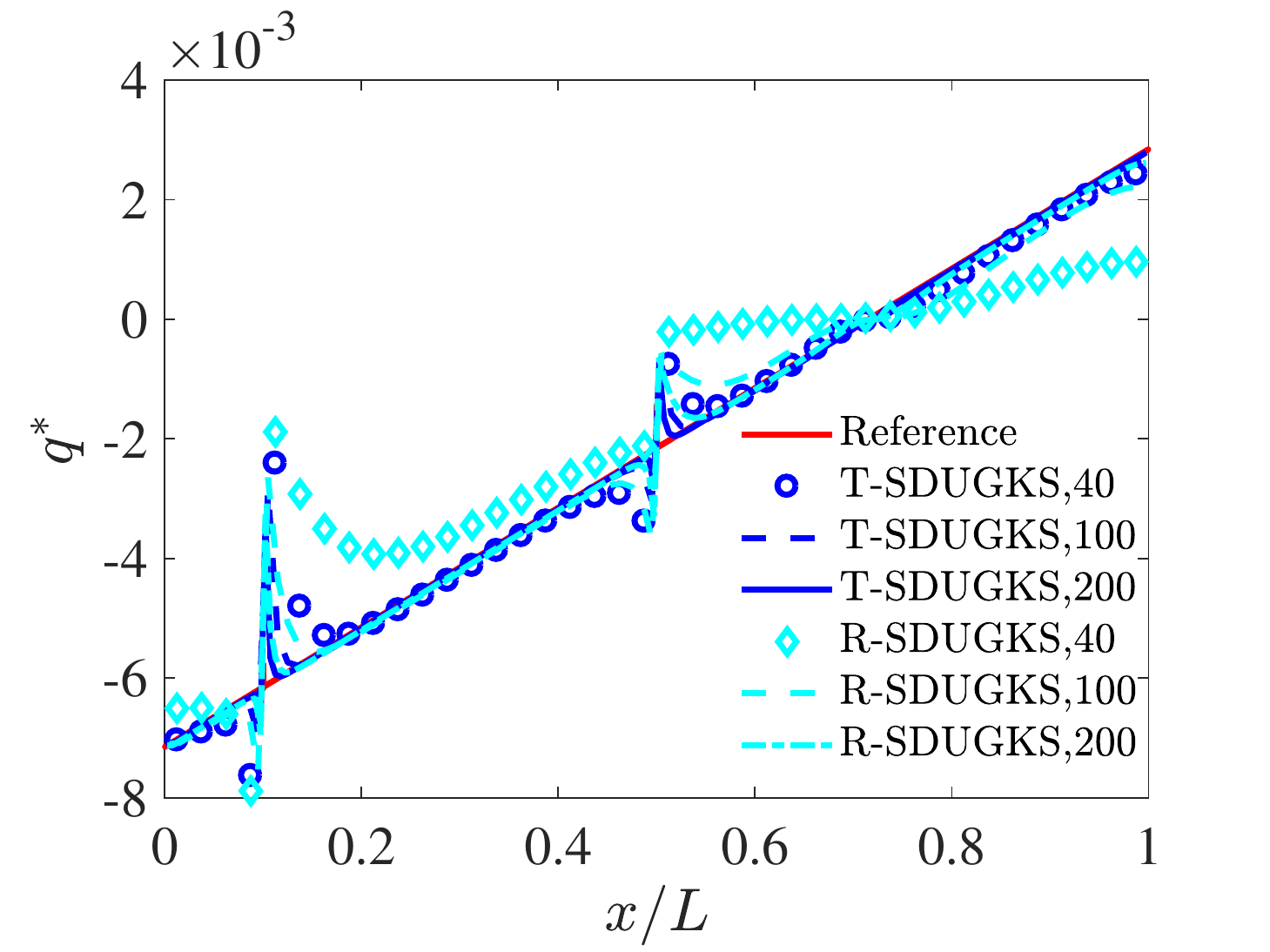}}~~
 \caption{Radiation energy and heat flux profiles of the T-SDUGKS and R-SDUGKS.
 }
 \label{figure34}
\end{figure}
The radiation energy and heat flux are plotted in Fig.~\ref{figure34}. It can be observed that the R-SDUGKS predicts almost incorrect results with the mesh of $N_x=40$ while the T-SDUGKS gives the reasonable ones. With the refinement of the mesh, the results become more accurate, and the T-SDUGKS can give better results than the R-SDUGKS under the same mesh.
These results clearly demonstrate that the T-SDUGKS is more accurate and robust than the R-SDUGKS for this problem.

\subsection{Radiation in a slab with large temperature gradient}
The nonequilibrium radiation transport in a slab with large temperature gradient is simulated by the T-SDUGKS and DD method to test their robustness.
\begin{figure}[!b]
 \centering
\subfloat[]{\includegraphics[width=0.35\textwidth]{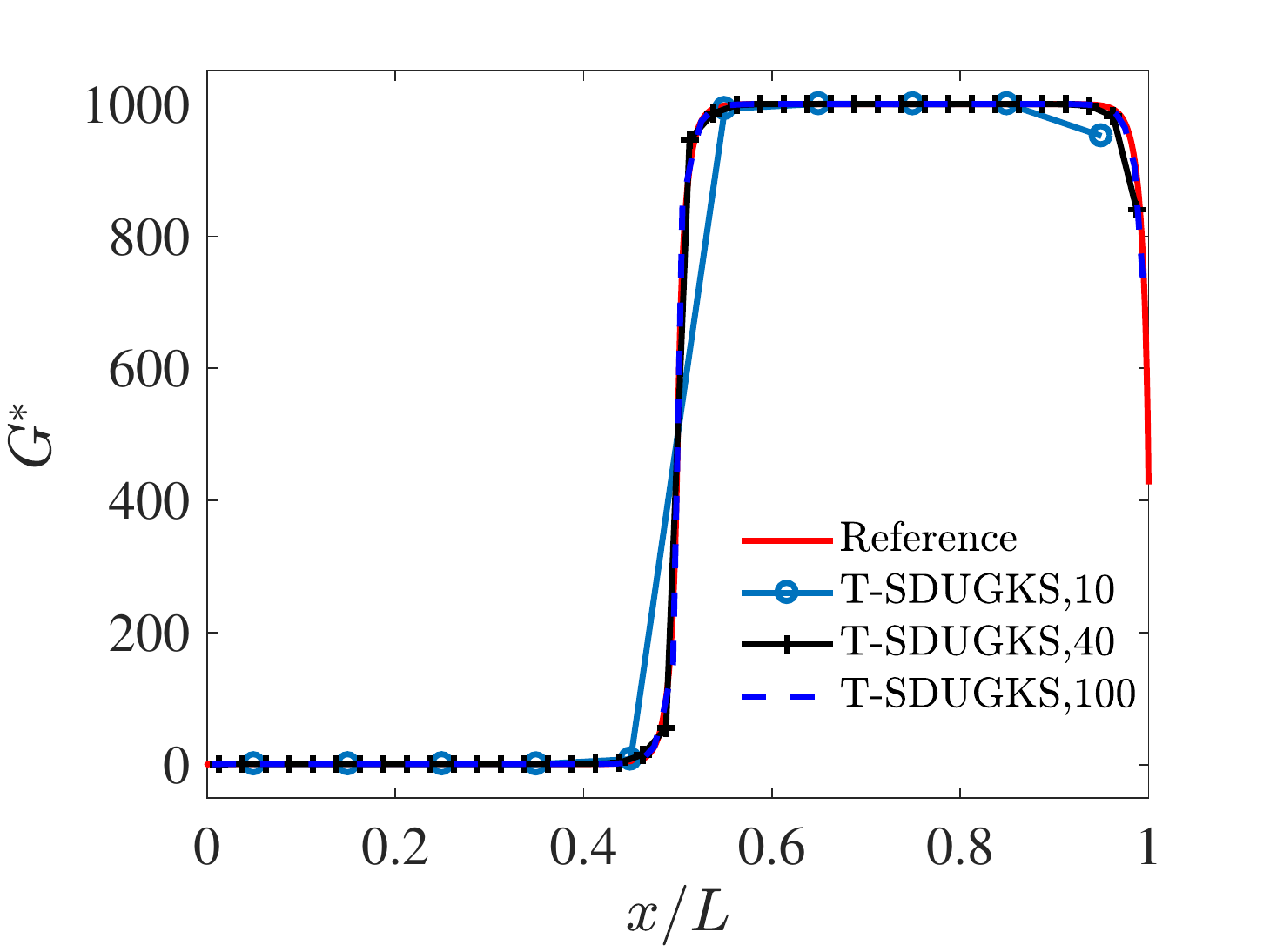}}~~
 \subfloat[]{\includegraphics[width=0.35\textwidth]{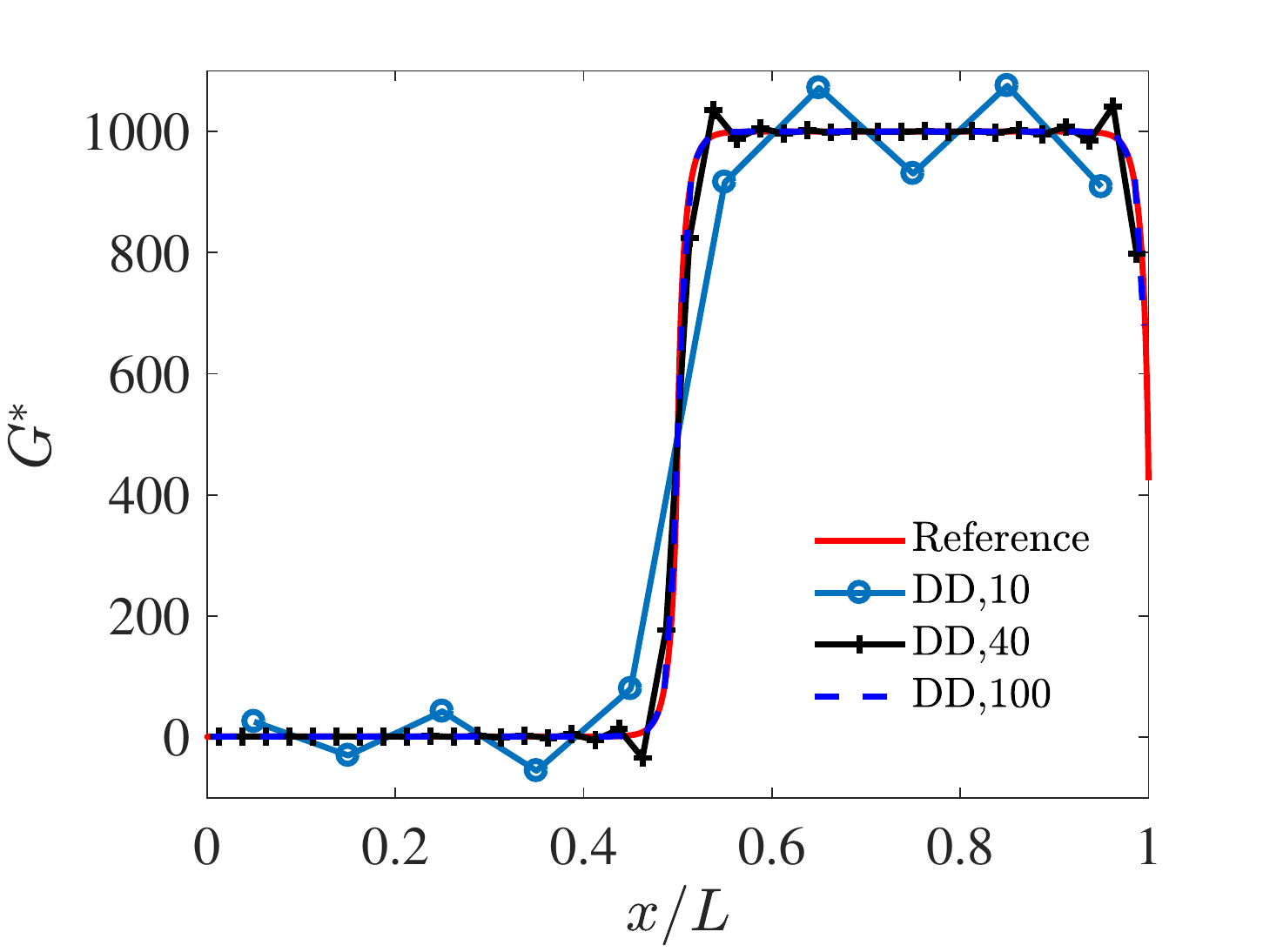}}~~\\
 \subfloat[]{\includegraphics[width=0.35\textwidth]{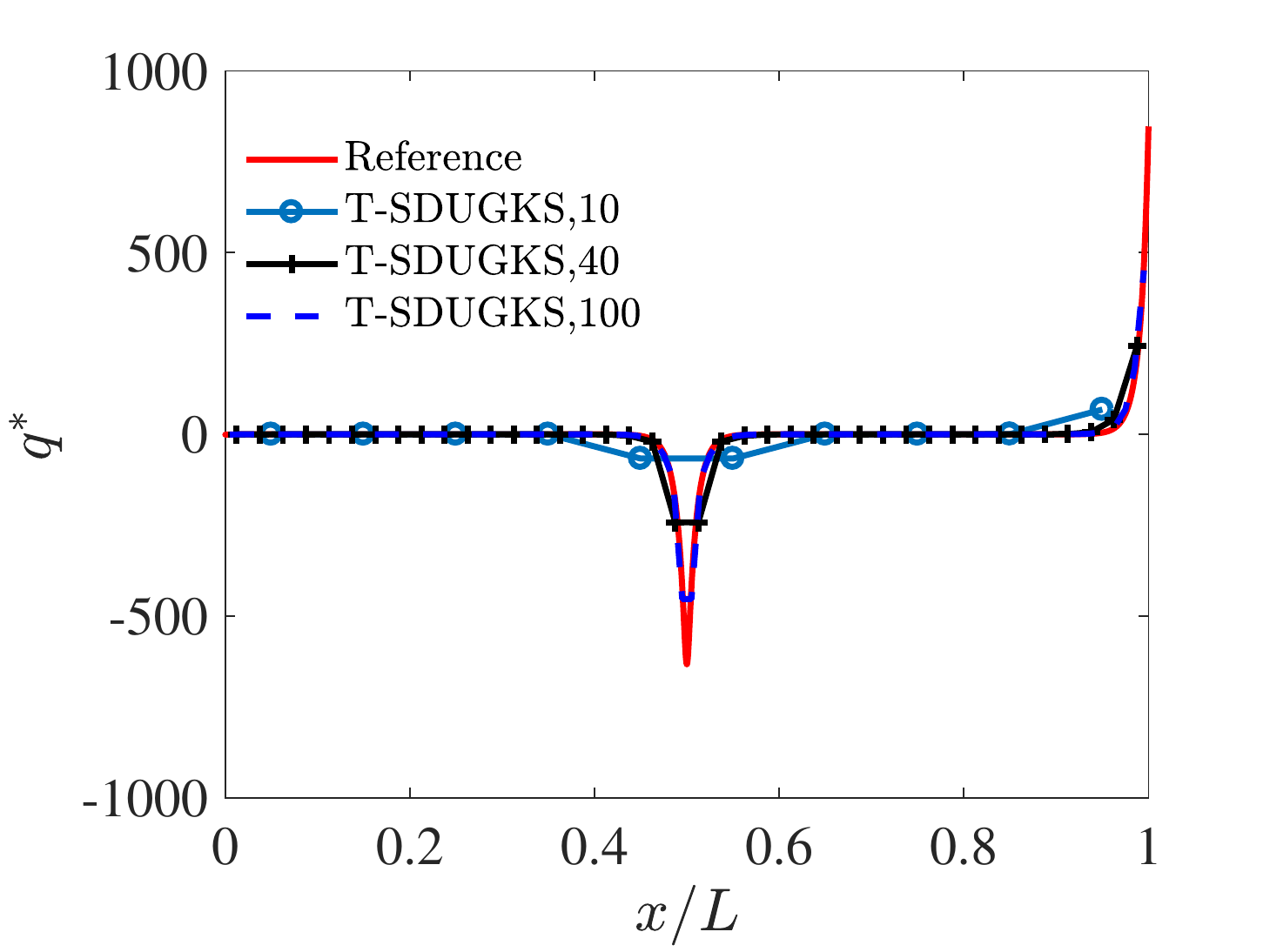}}~~
 \subfloat[]{\includegraphics[width=0.35\textwidth]{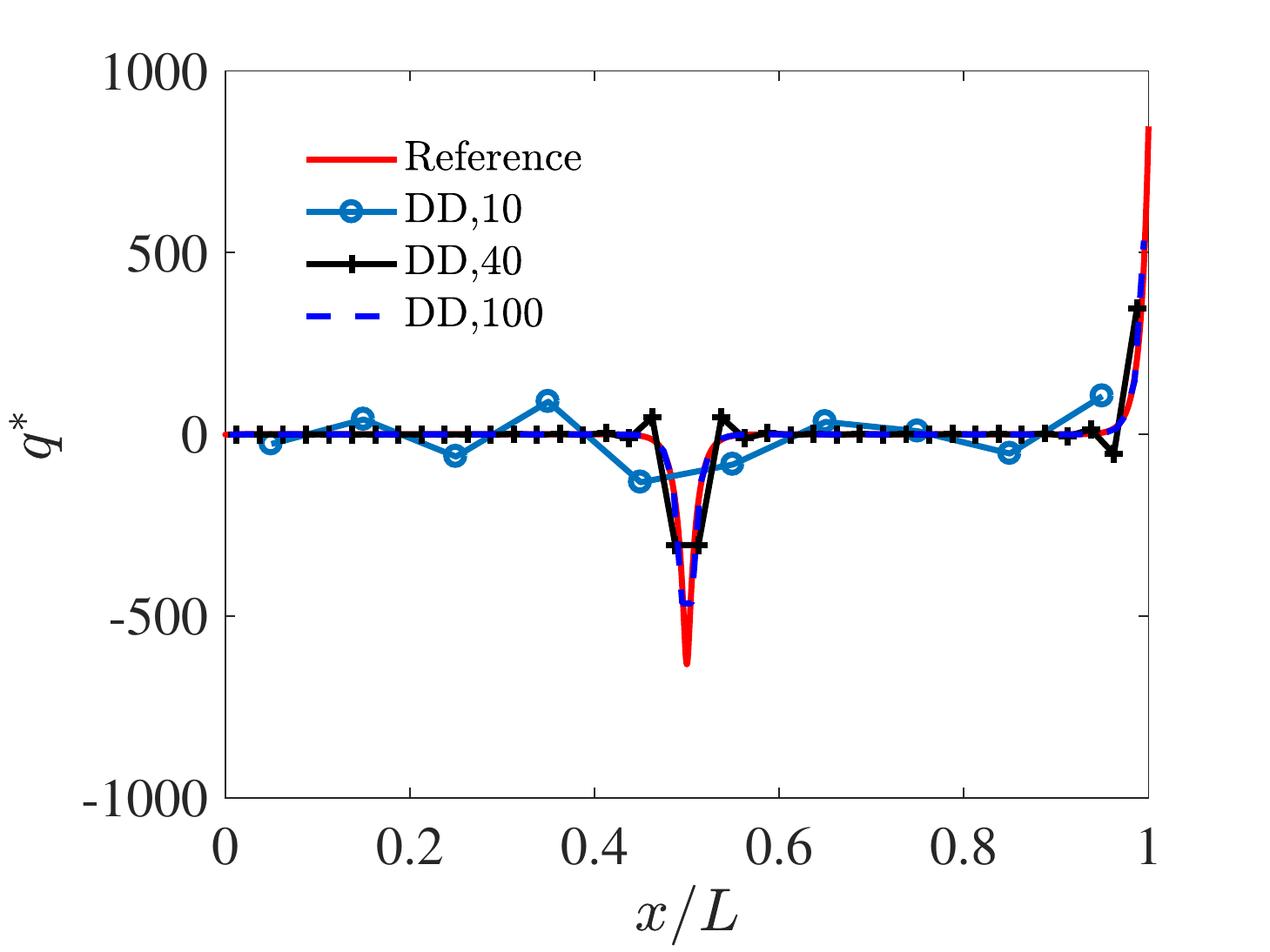}}~~
 \caption{Radiation energy and heat flux with T-SDUGKS and DD method at $\tau_{L}=100$.
 }
 \label{figure21}
\end{figure}%
Similar to the structure shown in Fig.~\ref{tutwozero}, both the left and right walls are black media and kept clod with $\Phi_{0} =0$, and the inside participating medium is kept hot with a large temperature gradient,
where the emissive power inside the slab is assumed to be
\begin{equation}
    \Phi\left( x \right)=\frac{\Phi_{max}+\Phi_{min}}{2}+\frac{\Phi_{max}-\Phi_{min}}{2} \tanh \left( \frac{x-L/2}{W} \right),
\label{eq:gradient}
\end{equation}
where $\Phi_{min}$ and $\Phi_{max}$ are the minimum and maximum values of  emissive power in the slab, respectively.
In the simulations, the scattering albedo is set to be $\omega =0.5$, the optical thickness is set to be $\tau_{L}=100,500$,
and  $\Phi_{min}=1$, $\Phi_{max}=1000$, $W=0.002$.
The smooth linear interpolation is used in T-SDUGKS for reconstructing the interface radiation intensity.
The angular space is discretized using the Gauss-Legendre quadrature with $N_{\mu}=100$, and the physical space is discretized into $N_{x}$ uniform cells. The DUGKS solution with 10000 uniform cells is taken as the reference solution.

Figure~\ref{figure21} shows the radiation energy and heat flux from   T-SDUGKS and DD method at $\tau_{L}=100$. It can be observed that the DD method produces nonphysical oscillations with mesh resolutions of $N_x=10$ and 40 which can be attributed to the unresolved mean free path $\left( \Delta x/\lambda>1 \right)$. On the other hand, the results given by T-SDUGKS are in  good agreement with the reference solutions even with the coarsest mesh $\left( N_x =10 \right)$ indicating that the mesh size  is not restricted by the  MFP.
\begin{figure}[!ht]
 \centering
\subfloat[]{\includegraphics[width=0.35\textwidth]{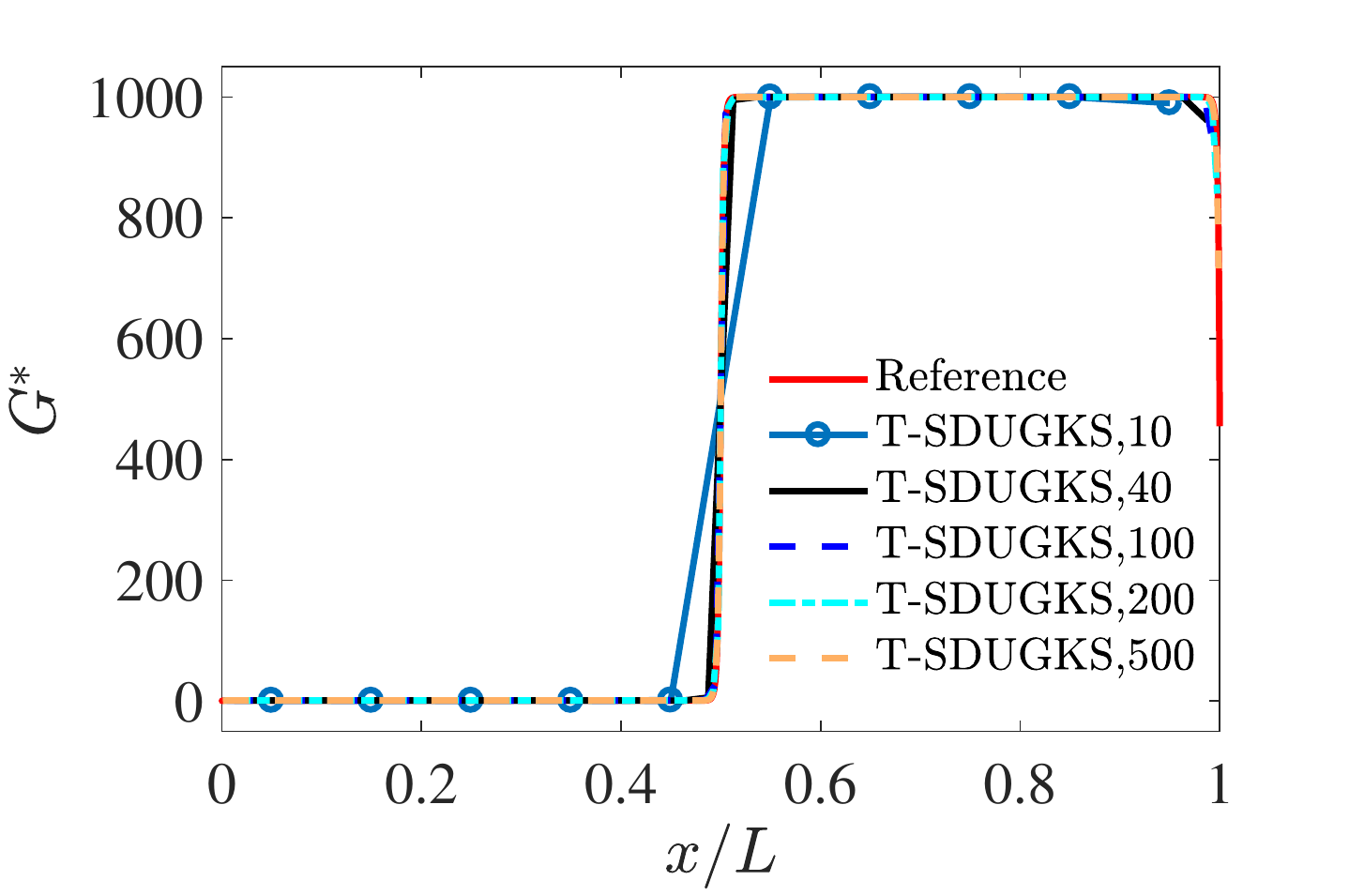}}~~
 \subfloat[]{\includegraphics[width=0.35\textwidth]{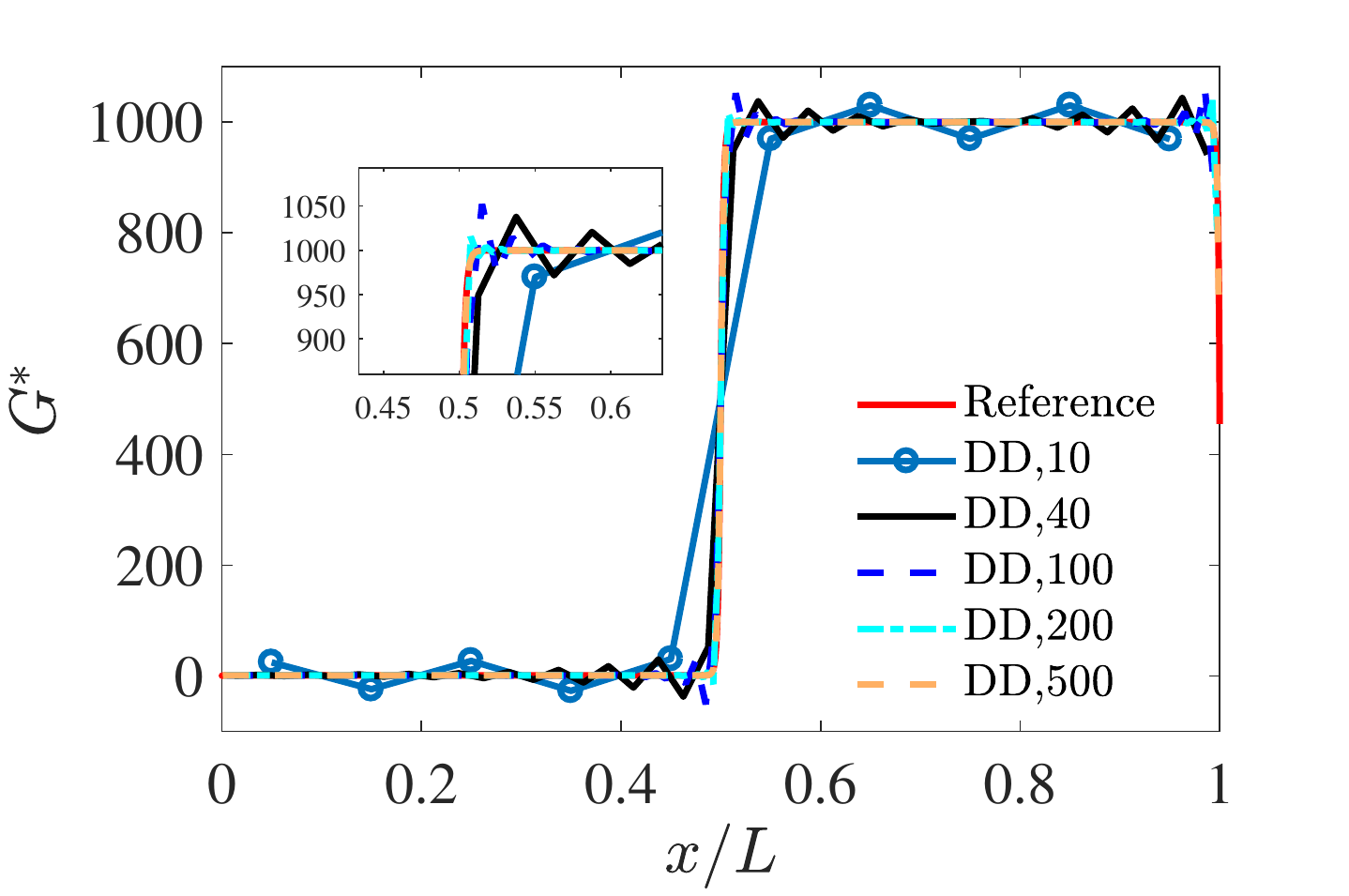}}~~\\
 \subfloat[]{\includegraphics[width=0.35\textwidth]{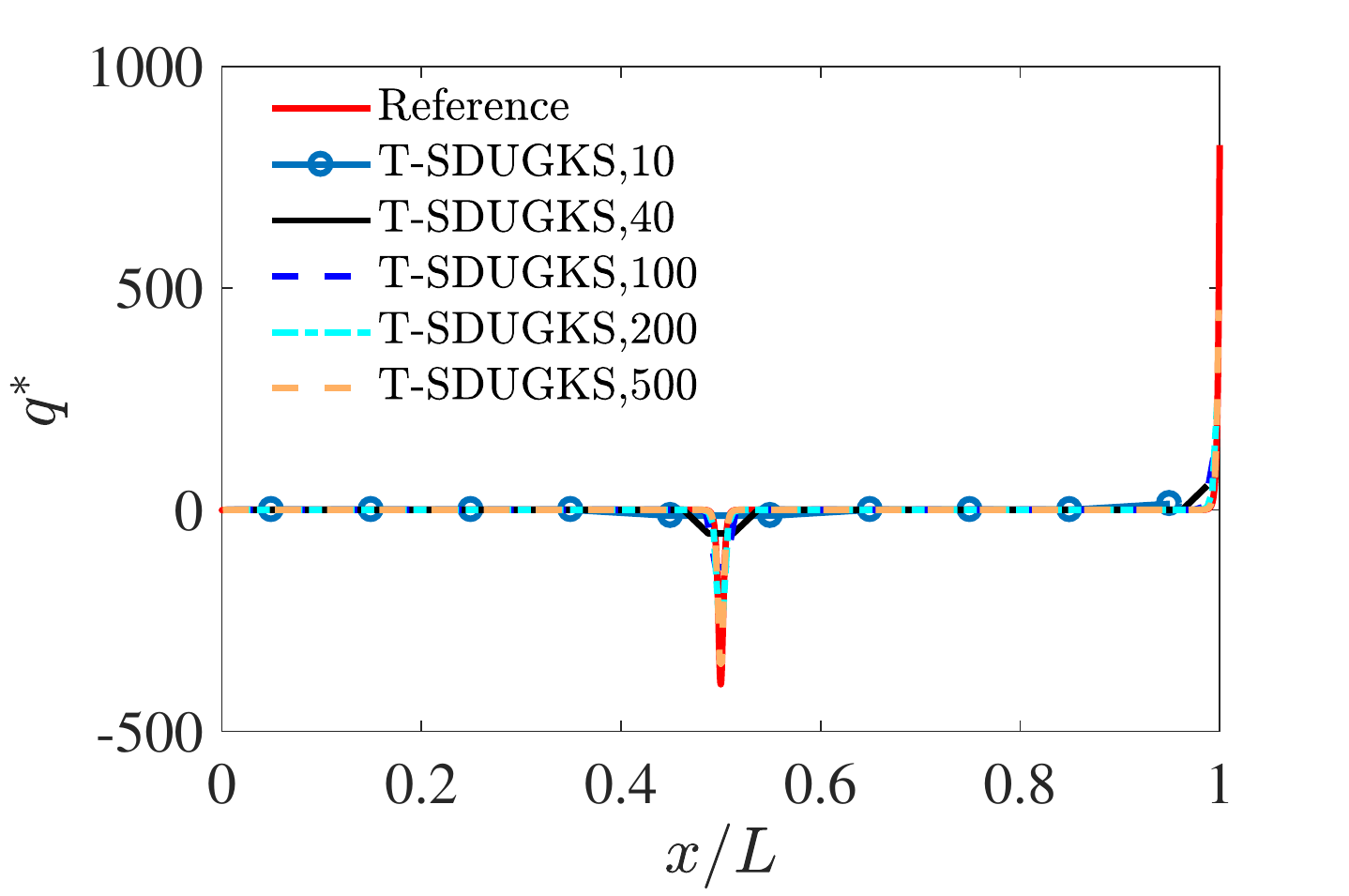}}~~
 \subfloat[]{\includegraphics[width=0.35\textwidth]{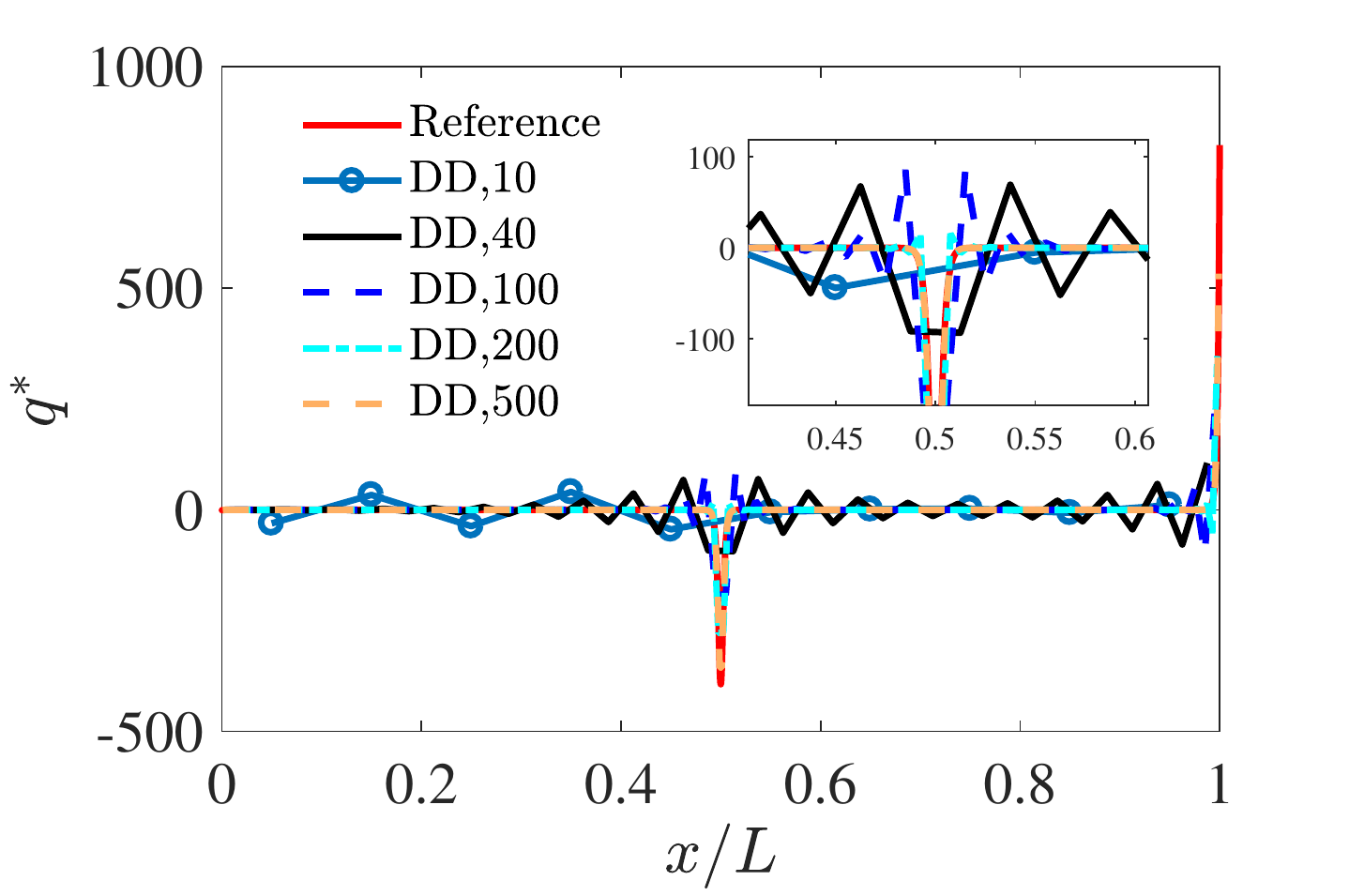}}~~
 \caption{Radiation energy and heat flux with T-SDUGKS and DD method at $\tau_{L}=500$.
 }
 \label{figure22}
\end{figure}
The results with $\tau_{L}=500$ are shown in Fig.~\ref{figure22}. It can be seen that as $\tau_{L}$ increases, the mesh resolution should be further increased for DD method to avoid nonphysical oscillations.
The above results demonstrate that the T-SDUGKS is more stable and robust than the DD method in solving radiative heat transfer problems with large temperature gradient and optical thickness.

\section{Conclusion}\label{sec5}
In this work, a  steady discrete unified gas kinetic scheme using trapezoidal rule in the reconstruction of interface radiation intensity is developed for multiscale radiative heat transfer problems.
With the coupling of the scattering, absorbing and emitting along the characteristic line of the RTE in determining the cell interface intensity, the T-SDUGKS can simulate multiscale radiative heat transfer problems without the limitation that the mesh size must be smaller than the photon MFP.
The T-SDUGKS is validated by the one-dimensional multiscale radiation problems in a slab and two-dimensional multiscale radiation problems in a black square enclosure.
Numerical results show that both the present T-SDUGKS and the original R-SDUGKS have faster convergence rates than the transient DUGKS for steady  problems, especially  in optical thin regions.
It is also found that the present T-SDUGKS is more accurate than the R-SDUGKS, and the characteristic line length is not limited by the extinction coefficient.
Furthermore, it is found that the mesh size in the T-SDUGKS is not restricted by the photon MFP, unlike the diamond difference method.
In summary, the proposed T-SDUGKS is an accurate, efficient and robust scheme for steady multiscale radiative heat transfer.

The proposed T-SDUGKS is designed based on the gray and isotropic RTE model, where the frequency dependence of the intensity and anisotropic of the scattering term are ignored.
The future work will extend the current method to problems with these effects.

\section*{Acknowledgments}\label{sec6}

This work was supported by the Fundamental Research Funds for the Central Universities (HUST: 2019kfyXMBZ040), the National Natural Science Foundation of China (No.12005073, No.12002131), and the China Postdoctoral Science Foundation (2021M701565).

\section*{DATA AVAILABILITY STATEMENT}
The data that support the findings of this study are available from the corresponding author upon reasonable request.

\bibliographystyle{IEEEtr}
\bibliography{Tx_SDUGKS}

\begin{thebibliography}{10}

\bibitem{machado2020experimental}
I.~M. Machado, P.~Pagot, and F.~M. Pereira, ``Experimental study of radiative
  heat transfer from laminar non-premixed methane flames diluted with co2 and
  n2,'' {\em International Journal of Heat and Mass Transfer}, vol.~158,
  p.~119984, 2020.

\bibitem{gunnarsson2020full}
A.~Gunnarsson, K.~Andersson, B.~R. Adams, and C.~Fredriksson, ``Full-scale
  3d-modelling of the radiative heat transfer in rotary kilns with a present
  bed material,'' {\em International Journal of Heat and Mass Transfer},
  vol.~147, p.~118924, 2020.

\bibitem{hunter2014normalization}
B.~Hunter and Z.~Guo, ``Normalization of various phase functions for radiative
  heat transfer analysis in a solar absorber tube,'' {\em Heat transfer
  engineering}, vol.~35, no.~6-8, pp.~791--801, 2014.

\bibitem{fuqiang2017radiative}
W.~Fuqiang, M.~Lanxin, C.~Ziming, T.~Jianyu, H.~Xing, and L.~Linhua,
  ``Radiative heat transfer in solar thermochemical particle reactor: a
  comprehensive review,'' {\em Renewable and Sustainable Energy Reviews},
  vol.~73, pp.~935--949, 2017.

\bibitem{chahlafi2012radiative}
M.~Chahlafi, F.~Bellet, F.~Fichot, and J.~Taine, ``Radiative transfer within
  non beerian porous media with semitransparent and opaque phases in non
  equilibrium: Application to reflooding of a nuclear reactor,'' {\em
  International journal of heat and mass transfer}, vol.~55, no.~13-14,
  pp.~3666--3676, 2012.

\bibitem{liou2002introduction}
K.-N. Liou, {\em An introduction to atmospheric radiation}.
\newblock Elsevier, 2002.

\bibitem{huang2022effects}
J.~Huang, C.~Lin, Y.~Li, and B.~Huang, ``Effects of humidity, aerosol, and
  cloud on subambient radiative cooling,'' {\em International Journal of Heat
  and Mass Transfer}, vol.~186, p.~122438, 2022.

\bibitem{broc1998nonequilibrium}
A.~Broc, V.~Joly, J.-P. Lafon, and C.~Marmignon, ``Nonequilibrium radiative
  hypersonic flows: aerospace applications,'' {\em Astrophysics and space
  science}, vol.~260, no.~1, pp.~29--43, 1998.

\bibitem{shang2012nonequilibrium}
J.~Shang and S.~Surzhikov, ``Nonequilibrium radiative hypersonic flow
  simulation,'' {\em Progress in Aerospace Sciences}, vol.~53, pp.~46--65,
  2012.

\bibitem{gorpas2010three}
D.~Gorpas, D.~Yova, and K.~Politopoulos, ``A three-dimensional finite elements
  approach for the coupled radiative transfer equation and diffusion
  approximation modeling in fluorescence imaging,'' {\em Journal of
  Quantitative Spectroscopy and Radiative Transfer}, vol.~111, no.~4,
  pp.~553--568, 2010.

\bibitem{zhang2007nano}
Z.~M. Zhang, Z.~M. Zhang, and Luby, {\em Nano/microscale heat transfer},
  vol.~410.
\newblock Springer, 2007.

\bibitem{modest2013radiative}
M.~F. Modest, {\em Radiative heat transfer}.
\newblock Academic press, 2013.

\bibitem{roger2014hybrid}
M.~Roger, C.~Caliot, N.~Crouseilles, and P.~Coelho, ``A hybrid
  transport-diffusion model for radiative transfer in absorbing and scattering
  media,'' {\em Journal of Computational Physics}, vol.~275, pp.~346--362,
  2014.

\bibitem{howell1969application}
J.~R. Howell, ``Application of monte carlo to heat transfer problems,'' in {\em
  Advances in heat transfer}, vol.~5, pp.~1--54, Elsevier, 1969.

\bibitem{rubinstein2016simulation}
R.~Y. Rubinstein and D.~P. Kroese, {\em Simulation and the Monte Carlo method}.
\newblock John Wiley \& Sons, 2016.

\bibitem{truelove1987discrete}
Truelove and S.~J., ``Discrete-ordinate solutions of the radiation transport
  equation,'' {\em Journal of Heat Transfer}, vol.~109:4, no.~4,
  pp.~1048--1051, 1987.

\bibitem{zhou2019modified}
R.-R. Zhou and B.-W. Li, ``The modified discrete ordinates method for radiative
  heat transfer in two-dimensional cylindrical medium,'' {\em International
  Journal of Heat and Mass Transfer}, vol.~139, pp.~1018--1030, 2019.

\bibitem{grissa2007three}
H.~Grissa, F.~Askri, M.~B. Salah, and S.~B. Nasrallah, ``Three-dimensional
  radiative transfer modeling using the control volume finite element method,''
  {\em Journal of Quantitative Spectroscopy and Radiative Transfer}, vol.~105,
  no.~3, pp.~388--404, 2007.

\bibitem{raithby1990finite}
G.~D. Raithby and E.~H. Chui, ``A finite-volume method for predicting a radiant
  heat transfer in enclosures with participating media,'' {\em Asme
  Transactions Journal of Heat Transfer}, vol.~112, no.~2, 1990.

\bibitem{raithby1999discussion}
G.~Raithby, ``Discussion of the finite-volume method for radiation, and its
  application using 3d unstructured meshes,'' {\em Numerical Heat Transfer:
  Part B: Fundamentals}, vol.~35, no.~4, pp.~389--405, 1999.

\bibitem{zhao2007solution}
J.~Zhao and L.~Liu, ``Solution of radiative heat transfer in graded index media
  by least square spectral element method,'' {\em International Journal of Heat
  and Mass Transfer}, vol.~50, no.~13-14, pp.~2634--2642, 2007.

\bibitem{hottel1958radiant}
H.~Hottel and E.~Cohen, ``Radiant heat exchange in a gas-filled enclosure:
  Allowance for nonuniformity of gas temperature,'' {\em AIChE Journal},
  vol.~4, no.~1, pp.~3--14, 1958.

\bibitem{henson1997comparison}
J.~C. Henson and W.~Malalasekera, ``Comparison of the discrete transfer and
  monte carlo methods for radiative heat transfer in three-dimensional
  nonhomogeneous scattering media,'' {\em Numerical Heat Transfer, Part A
  Applications}, vol.~32, no.~1, pp.~19--36, 1997.

\bibitem{sarvari2017solution}
S.~H. Sarvari, ``Solution of multi-dimensional radiative heat transfer in
  graded index media using the discrete transfer method,'' {\em International
  Journal of Heat and Mass Transfer}, vol.~112, pp.~1098--1112, 2017.

\bibitem{zhou2004new}
H.-C. Zhou, D.-L. Chen, and Q.~Cheng, ``A new way to calculate radiative
  intensity and solve radiative transfer equation through using the monte carlo
  method,'' {\em Journal of Quantitative Spectroscopy and Radiative Transfer},
  vol.~83, no.~3-4, pp.~459--481, 2004.

\bibitem{huang2013solution}
Z.~Huang, H.~Zhou, Q.~Cheng, and P.-f. Hsu, ``Solution of radiative intensity
  with high directional resolution in three-dimensional rectangular enclosures
  by dresor method,'' {\em International Journal of Heat and Mass Transfer},
  vol.~60, pp.~81--87, 2013.

\bibitem{kulacki2018handbook}
F.~A. Kulacki, S.~Acharya, Y.~Chudnovsky, R.~M. Cotta, R.~Devireddy, V.~K.
  Dhir, M.~P. Meng{\"u}{\c{c}}, J.~Mostaghimi, and K.~Vafai, {\em Handbook of
  thermal science and engineering}.
\newblock Springer, 2018.

\bibitem{howell2020thermal}
J.~R. Howell, M.~P. Meng{\"u}{\c{c}}, K.~Daun, and R.~Siegel, {\em Thermal
  radiation heat transfer}.
\newblock CRC press, 2020.

\bibitem{wang2019quantitative}
D.~Wang and H.~Zhou, ``Quantitative evaluation of the computational accuracy
  for the monte carlo calculation of radiative heat transfer,'' {\em Journal of
  Quantitative Spectroscopy and Radiative Transfer}, vol.~226, pp.~100--114,
  2019.

\bibitem{golse2003domain}
F.~Golse, S.~Jin, and C.~D. Levermore, ``A domain decomposition analysis for a
  two-scale linear transport problem,'' {\em ESAIM: Mathematical Modelling and
  Numerical Analysis}, vol.~37, no.~6, pp.~869--892, 2003.

\bibitem{roger2013dynamic}
M.~Roger and N.~Crouseilles, ``A dynamic multi-scale model for transient
  radiative transfer calculations,'' {\em Journal of Quantitative Spectroscopy
  and Radiative Transfer}, vol.~116, pp.~110--121, 2013.

\bibitem{coelho2016multi}
P.~J. Coelho, N.~Crouseilles, P.~Pereira, and M.~Roger, ``Multi-scale methods
  for the solution of the radiative transfer equation,'' {\em Journal of
  Quantitative Spectroscopy and Radiative Transfer}, vol.~172, pp.~36--49,
  2016.

\bibitem{xu2010unified}
K.~Xu and J.-C. Huang, ``A unified gas-kinetic scheme for continuum and
  rarefied flows,'' {\em Journal of Computational Physics}, vol.~229, no.~20,
  pp.~7747--7764, 2010.

\bibitem{zhu2021first}
Y.~Zhu and K.~Xu, ``The first decade of unified gas kinetic scheme,'' {\em
  arXiv preprint arXiv:2102.01261}, 2021.

\bibitem{guo2013discrete}
Z.~Guo, K.~Xu, and R.~Wang, ``Discrete unified gas kinetic scheme for all
  knudsen number flows: Low-speed isothermal case,'' {\em Physical Review E},
  vol.~88, no.~3, p.~033305, 2013.

\bibitem{guo2021progress}
Z.~Guo and K.~Xu, ``Progress of discrete unified gas-kinetic scheme for
  multiscale flows,'' {\em Advances in Aerodynamics}, vol.~3, no.~1, pp.~1--42,
  2021.

\bibitem{mieussens2013asymptotic}
L.~Mieussens, ``On the asymptotic preserving property of the unified gas
  kinetic scheme for the diffusion limit of linear kinetic models,'' {\em
  Journal of Computational Physics}, vol.~253, pp.~138--156, 2013.

\bibitem{sun2015asymptotic}
W.~Sun, S.~Jiang, and K.~Xu, ``An asymptotic preserving unified gas kinetic
  scheme for gray radiative transfer equations,'' {\em Journal of Computational
  Physics}, vol.~285, pp.~265--279, 2015.

\bibitem{luo2018multiscale}
X.-P. Luo, C.-H. Wang, Y.~Zhang, H.-L. Yi, and H.-P. Tan, ``Multiscale
  solutions of radiative heat transfer by the discrete unified gas kinetic
  scheme,'' {\em Physical Review E}, vol.~97, no.~6, p.~063302, 2018.

\bibitem{song2020discrete}
X.~Song, C.~Zhang, X.~Zhou, and Z.~Guo, ``Discrete unified gas kinetic scheme
  for multiscale anisotropic radiative heat transfer,'' {\em Advances in
  Aerodynamics}, vol.~2, no.~1, pp.~1--15, 2020.

\bibitem{shi2021improved}
Y.~Shi, W.~Sun, L.~Li, and P.~Song, ``An improved unified gas kinetic particle
  method for radiative transfer equations,'' {\em Journal of Quantitative
  Spectroscopy and Radiative Transfer}, vol.~261, p.~107428, 2021.

\bibitem{zhou2020discrete}
X.~Zhou and Z.~Guo, ``Discrete unified gas kinetic scheme for steady multiscale
  neutron transport,'' {\em Journal of Computational Physics}, vol.~423,
  p.~109767, 2020.

\bibitem{van1977towards}
B.~Van~Leer, ``Towards the ultimate conservative difference scheme. iv. a new
  approach to numerical convection,'' {\em Journal of computational physics},
  vol.~23, no.~3, pp.~276--299, 1977.

\bibitem{abramowitz1972handbook}
M.~Abramowitz and I.~A. Stegun, {\em Handbook of mathematical functions with
  formulas, graphs, and mathematical tables}, vol.~55.
\newblock US Government printing office, 1972.

\bibitem{heaslet1965radiative}
M.~A. Heaslet and R.~F. Warming, ``Radiative transport and wall temperature
  slip in an absorbing planar medium,'' {\em International Journal of Heat and
  Mass Transfer}, vol.~8, no.~7, pp.~979--994, 1965.

\bibitem{adams2002fast}
M.~L. Adams and E.~W. Larsen, ``Fast iterative methods for discrete-ordinates
  particle transport calculations,'' {\em Progress in nuclear energy}, vol.~40,
  no.~1, pp.~3--159, 2002.

\end{thebibliography}

\end{document}